\documentclass[12pt]{iopart}
\usepackage{pstricks}%
\usepackage{graphicx}
\usepackage{dcolumn}
\usepackage{bm}
\usepackage{iopams}
\usepackage{pifont}
\usepackage{psfrag}
\usepackage[latin2]{inputenc}
\usepackage{oldgerm}
\usepackage{cite}

\newcommand\ket[1]{|{#1}\rangle}

\newcommand{\be}{\begin{equation}}
\newcommand{\ee}{\end{equation}}

\newcommand{\sumlim}{\sum\limits}

\newcommand{\skipc}[2]{}

\newcommand{\fig}[1]{Figure~\ref{#1}}
\newcommand{\eq}[1]{(\ref{#1})}

\begin{document}

\title[Factorization  with Gauss sums: Mathematical background]{Factorization of numbers with Gau{ss} sums:
I. Mathematical background}

\author{S W\"olk$^1$, W Merkel$^1$, W P Schleich$^1$, I Sh Averbukh$^2$ and 
B Girard$^3$ }

\address{$^1$ Institut f\"ur Quantenphysik, Universit\"at Ulm,\\ Albert-Einstein-Allee 11, D-89081 Ulm, Germany 
}%

\address{$^2$ Department of Chemical Physics,\\ Weizmann Institute of Science, Rehovot 76100, Israel 
}

\address{$^3$ Laboratoire de Collisions, Agr\'egats, R\'eactivit\'e,   IRSAMC (Universit\'e de Toulouse/UPS; CNRS) Toulouse, France
}

\ead{sabine.woelk@uni-ulm.de}

\date{\today}

\begin{abstract}

We use the periodicity properties of  generalized Gau{ss} sums to factor numbers. Moreover, we derive rules for finding the factors  and illustrate this factorization scheme for various 
examples. This algorithm relies solely on interference and scales exponentially. 
\end{abstract}


\submitto{New J. of Physics}

\section{Introduction}
\label{gauss:sums:physics}

"Mathematics is the abstract key that turns the lock of the physical universe." This quote by John Polkinghorne expresses in  a poetic way the fact that 
physics  uses mathematics as a tool to make predictions about physical phenomena such as the motion of a particle, the outcome of a measurement, or the time evolution of a quantum state. However, there exist situations in which the roles of the two disciplines are interchanged and physical phenomena allow us  to obtain mathematical quantities. In the present series of papers \cite{merkel:2010} we follow this path   of employing nature to evaluate mathematical functions. Here, we  choose the special example of Gauss sums \cite{lang:1970} and show that they are ideal for factoring numbers. At the same time we must  issue the caveat that  in constrast to the Shor algorithm  \cite{shor:1994}, the proposed  Gauss sum factorization algorithm in its most elementary version  scales only exponentially since it is based solely on interference and does not involve entanglement. However, there exist already indications \cite{woelk2011} that a combination of entanglement and Gauss sums can lead to a powerful tool to attack questions of factorization.

The recent years have seen an impressive number of experiments implementing Gauss sums in physical systems to factor numbers. These systems range from NMR methods \cite{mehring:2007,mahesh:2007,peng:2008} via cold atoms \cite{gilowski:2008} and Bose-Einstein condensates \cite{sadgrove:2008,sadgrove:2009}, tailored ultra-short laser pulses \cite{bigourd:2008,weber:2008} to classical light in a multi-path  Michelson interferometer \cite{tamma:2009,tamma:2009:b}. Although these experiments have been motivated by earlier versions of this series of papers made available before publication, they have focused exclusively on a very special type of Gauss sum, that is the truncated Gauss sum, which is not at the center of the present work. Indeed, throughout this series we concentrate on three types of Gauss sums: the continuous, the discrete  and the reciprocate Gauss sum. Moreover, we propose experimental realizations with the help of chirped laser pulses \cite{girard:2003,girard:2004} interacting with appropriate atoms.

We persue in two steps: ({\it i}) We first show that  the mathematical properties of  Gau{ss} sums allow us to factor numbers, and (ii) we then present three elementary quantum systems to implement our method using chirped laser pulses and multi-level atoms. To each step we dedicate an article.There exists a good reason to separate the individual articles. The work presented in the two articles is a combination of quantum optics and number theory \cite{number_theory}. In order to avoid overloading the quantum optical aspects of the problem with number-theoretical questions we deal with the mathematical properties of Gau{ss} sums in the present  article and adress the realizations in the second part.

\subsection{Gauss sums in physics}
Gauss sums \cite{davenport:1980,schleich:2005:primes} manifest themselves in many phenomena in physics and come in different varieties. They are similar to Fourier sums with the distinct difference that the summation index appears in the phase in a quadratic rather than a linear way.

Real-valued Gaussians are familiar from statistical physics. The integral over a finite extension of a Gaussian leads to a higher transcendental, that is the error function. When the integration is not along the real axis but along one of the diagonals of complex space we arrive at an integral giving rise to the Cornu spiral. This geometrical object determines \cite{born} the intensity distribution of light on a screen in the far field of a edge. It is also the essential building block of the Feynman path integral \cite{feynman:1965,wheeler:1989}and the method of stationary phase \cite{dowling:1991}.

A modern application of integrals over quadratic phases in physics are chirped laser pulses. Here, the frequency of the light increases linearly in time giving rise to a quadratic phase. The resulting excitation probability is determined \cite{merkel:2007} by the complex-valued error function which is closely related to the Cornu spiral. Indeed, the Cornu spiral was
observed in real-time excitation of Rb atoms by chirped pulses \cite{zamith:2001}.

A generalization of the Cornu spiral arises when we replace the integration over the continuous variable by a summation over  a discrete parameter. In this case the Cornu spiral turns into a Gauss sum. 

Sums have properties which are dramatically different from their corresponding integrals. This feature stands out most clearly in the phenomena of revivals and fractional revivals \cite{leichtle:PRA:1996,schleich:2001:thebook}. In the context of the Cornu spiral the transition from the integral to the sum leads to the curlicues \cite{berry:1988:a,berry:1988:b}, that is Cornu spirals in Cornu spirals. The origin of this self-similarity is discreteness. In the Gauss sum the question if a parameter takes on integer, rational or irrational values is crucial whereas in the error function integral it does not matter. This feature will also be of importance in the context of the factorization of numbers which is the topic of the present series of articles.

Gauss sums are at the very heart of the Talbot effect \cite{arndt:2009} which describes the intensity distribution of the light close to a diffraction grating. They are also crucial in the familiar problem of the particle in the box and are the origin of the design of quantum carpets \cite{berry:2001} and the creation of superpositions of distinct phase states \cite{schleich:1993} by non-linear evolution \cite{sanders:1992}. Moreover, there is an interesting connection \cite{schleich:2005:primes} to the Riemann zeta function which is determined by the Mellin transform of the Jacobi theta function. For complex-valued rational arguments the zeta function leads to Gauss sums. 


\subsection{Why factor numbers with Gauss sums?\label{ssec:why}}
It is the exponential increase of the dimension of Hilbert space together with entanglement which makes the quantum computer so efficient and ideal for the problem of factoring. Two landmark experiments \cite{vandersypen:2001,lanyon:2007} have implemented the Shor algorithm. They were based either on Nuclear Magnetic Resonance (NMR) \cite{vandersypen:2001}, or optical \cite{lanyon:2007} techniques and were able to factor the number $15$.

Our proposal outlined in this series of articles follows a different approach. It aims for an analogue computer that relies solely on interference \cite{zubairy:2007} without using entanglement. We test if a given integer $\ell$ is a factor of the integer $N$, to be factored. Since we have to try out at least $\sqrt{N}$ such factors the method scales exponentially with the number of digits of $N$. It is therefore a classical algorithm. This fact is not surprising since our method does not take advantage  of the exponential resources of the Hilbert space.

At the heart of the Shor algorithm is the task of finding the period of a function \cite{mermin:2007}. In our approach we also use properties of a periodic function. It is the periodicity of the Gauss sum which allows us to factor numbers. For this purpose we construct a system whose output is a Gauss sum. In this sense the system, that is Nature, is performing the calculation for us. 

But why are Gauss sums ideal tools for the problem of factorization?  Gauss sums play an important role in number theory as well as physics. Three examples may  testify to this claim: (i) The distribution of prime numbers is determined by the non-trivial zeros of the Riemann zeta function which is related \cite{schleich:2005:primes} to the Gauss sums, (ii) a quantum algorithm \cite{vandam1,vandam2, vandam3} allows to calculate the phase of a Gauss sum efficiently, and (iii) the Gauss reciprocity law can be interpreted \cite{onishi:2003} as  the analogue of the commutation relation between position and momentum operators in quantum mechanics.

\subsection{Various classes of Gauss sums}

In order to lay the ground work for part II of this series we dedicate the present article to an overview of the mathematical properties of several Gau{ss} sums which arise in different physical systems. When we study the two-photon excitation through an equidistant ladder system with a chirped laser pulse we arrive \cite{girard:2004} at the sum
\be
S_N(\xi)\sim
\sum\limits_{m} w_m
\exp\left[2\pi \rmi\left(m+\frac{m^2}{N}\right)\xi\right].
\label{ex:prob:p2}
\ee
Here $N$ is expressed in terms of parameters characterizing the harmonic manifold and the argument $\xi$ is proportional to the rescaled chirp of the laser pulse. The distribution of weight factors $w_m$ is governed by the shape of the laser pulse. The sum $S_N$ contains linear as well as quadratic phases.  

Quite different sums emerge for a one-photon transition in a two-level atom interacting with two driving fields. Here we find two different types of Gauss sums depending on the temporal shape of the fields. For a sinusoidal modulation of the excited state and a weak driving field the corresponding excitation probability amplitude is proportional to the sum
\be
S_N(\ell)\sim
\sum\limits_m w_m 
\exp\left[2\pi \rmi\,m^2\frac{\ell}{N}\right]
\label{ex:prob:p3:1}
\ee
which depends on purely quadratic phases.

In the second realization the modulating field causes a linear variation of the excited state energy and the one-photon transition is driven by a train of equidistant delta-shaped laser pulses. The excitation probability amplitude is proportional to the sum
\be
{\cal A}_{N}(\ell)\sim\sum\limits_m 
\exp\left[-2\pi \rmi\, m^2\frac{N}{\ell}\right]
\label{ex:prob:p3:2}
\ee
at integer arguments $\ell$ which is again of the form of a Gauss sum. In comparison to $S_N(\ell)$ now the roles of argument $\ell$ and the number $N$ to be factored are interchanged.

\subsection{Summary of experimental work}
Several experiments \cite{mehring:2007, mahesh:2007, gilowski:2008, bigourd:2008, weber:2008, peng:2008, tamma:2009, tamma:2009:b, sadgrove:2008, sadgrove:2009} have already successfully demonstrated factorization with the help of Gauss sums. We now briefly summarize them and highlight important features in Table \ref{table:1}.

\begin{table}
\caption{Summary of characteristic features of experiments factoring numbers using Gauss sums. All experiments are based on the Gauss sum ${\cal A}_N^{(M)}(\ell)$ defined by \eq{gauss2.2}, or closely related ones. Here we list the system, the amount of digits of the number $N$ to be factored , the number of terms $M$ in the sum  and important features associated with the specific approach.  \label{table:1}}
\begin{indented}
 \item[]
\begin{tabular}{@{}lccl}
\br

System & $N$ & $M$ &\\ &(digits)& \\ \mr
NMR \cite{mehring:2007,mahesh:2007,peng:2008} & 17 & 10  & higher order \\ &&& truncated Gauss sum,\\ &&& random terms \\ 
cold atoms \cite{gilowski:2008} & 6 & 15  & large number of\\&&& pulses  \\ 
BEC \cite{sadgrove:2008,sadgrove:2009} & 5 & 10  & enhanced visibility\\ 
ultra short laser pulses \cite{bigourd:2008,weber:2008}&13& 30 & random terms \\ 
multi-path Michelson interferometer \cite{tamma:2009,tamma:2009:b} & 6&  3 &$N,\;\ell$ are varied \\&&&  \\
\br
\end{tabular}
\end{indented}

\end{table}

The first set of experiments relies on the interaction of a sequence of electromagnetic  pulses with a two-level quantum system. The phase of each  pulse is adjusted such that the total excitation probability is determined by a Gauss sum \cite{mehring:2007,stefanak:2008}. Two realizations of two-level systems have been pursued: in NMR experiments  \cite{mehring:2007,mahesh:2007,peng:2008} the two levels correspond to the two orientations of the spin, for example of the proton of the water molecule. However, also optical pumping in laser-cooled atoms creates \cite{gilowski:2008} an effective two-level situation, for example of the hyperfine states of rubidium.

The second type of experiments utilizes a sequence of appropriately designed femto-second laser pulses \cite{bigourd:2008,weber:2008}. The intensity at a given frequency component of this light is given by the interference of this component of the individual pulses giving rise to a sum. With the help of a pulse shaper it is possible to imprint the  phases necessary to obtain a Gauss sum determining the light intensity at a given frequency.

The third class of experiments \cite{tamma:2009,tamma:2009:b} is based on a multi-path Michelson interferometer \cite{rangelov:2009}. Here the phase shifts accumulated in the individual arms increase quadratically with the arm length. This experiment  does not suffer from the problem \cite{jones:2008}, that the ratio $N/\ell$ has to be precalculated. Moreover, they do not only calculate the truncated Gauss sum for integer trial factors $\ell$ but also for rational arguments $\xi=q/r$. Gauss sums at rational arguments open up a new avenue towards factorization \cite{woelk:2009}.

The experiment reported in Refs.\cite{sadgrove:2008,sadgrove:2009} plays a very special role in this gallery of factorization experiments using Gauss sums. (i) It is the only one so far that has used a Bose-Einstein condensate (BEC), (ii) the read-out utilizes the momentum distribution, and (iii) it relies on a generalization of Gauss sums. Furthermore, they claim a better visibility by using the probability distribution for higher momentums.

All of theses studies have implemented the  Gauss sum 
\be
{\cal A}_N^{(M)}(\ell)\equiv\frac{1}{M+1}\sum\limits_{m=0}^M \exp\left(-2\pi \rmi\, m^2 \frac{N}{\ell}\right)
\label{gauss2.2}
\ee
which is  closely related to the ones discussed in the present article. Here $M+1$ is the number of interfering paths which could be the number of laser pulses, or the number of arms in the Michelson interferometer.

With this type of Gauss sum the idea of factorization is straight-forward: When $\ell$ is a factor of $N$, then $N/ \ell$ is an integer. As a result, the phase of each term in the Gauss sum ${\cal A}_N^{(M)}(\ell)$ is an integer multiple of $2\pi$ and due to constructive interference the value of the sum is unity. For a non-factor $\ell$ the ratio $N/ \ell$ is  a rational number and due to the quadratic variation of the phase  the individual terms interfere destructively. This feature leads to the unique criterion for distinguishing factors from non-factors: For factors the sum ${\cal A}_N^{(M)}(\ell)$ is unity, whereas for non-factors it is not.

However, a closer analysis \cite{stefanak:2007} reveals the fact that for a subset of test factors, which are not factors, the Gauss sum ${\cal A}_N^{(M)}(\ell)$ takes on values rather close to unity. It it therefore difficult to distinguish them from real factors. These test factors have been called \textit{ghost factors}. The amount of ghost factors increases as the number $N$ to be factored increases. 

Two techniques to suppress ghost factors offer themselves: (i) a Monte-Carlo evaluation \cite{weber:2008,peng:2008} of the complete Gauss sum 
\be
{\cal A}_N^{(\ell-1)}(\ell)\equiv\frac{1}{\ell}\sum\limits_{m=0}^{\ell-1} \exp\left(-2\pi \rmi\, m^2 \frac{N}{\ell}\right),
\ee
and (ii) the use  \cite{stefanak:2008:b} of exponential sums 
\be
{\cal A}_N^{(j,M)}(\ell)\equiv\frac{1}{M+1}\sum\limits_{m=0}^M \exp\left(-2\pi \rmi\, m^j \frac{N}{\ell}\right)
\ee
with $1 \leq j$.

Ghost factors originate from the fact that the phase of consecutive terms in a Gauss sum such as in \eq{gauss2.2} can increase very slowly. To take terms of ${\cal A}_N^{(M)}$ randomly allows the Gauss sum to interfere destructively rather rapidly. This technique has been demonstrated in experiments with femto-second pulses \cite{weber:2008} and a 13-digits number could be factored with  very few pulses. Also a NMR experiment \cite{peng:2008} followed this approach and factored a 17-digits number consisting of two prime numbers. 

It is interesting to note that the argument for the factorization with Gauss sum ${\cal A}_N^{(M)}(\ell)$ does not make use of the fact that the phase varies quadratically. It is correct for any phase of the form $m^j$, where $j$ is an integer. The advantage of powers larger than two is that in this case the cancellation of neighboring terms is faster. This technique has been applied successfully in NMR experiments \cite{peng:2008} for $j=5$ and a 17-digits number could be factored.

\subsection{Alternative proposals of Gauss sum factorization}
It is interesting to compare and contrast the Gauss sums and our proposals for factoring numbers to other suggestions along these lines. Here, we refer especially to the pioneering algorithm outlined in Ref.\cite{clauser:1996} based on the Talbot effect. This method uses the intensity distribution of light in the near-field of a $N$-slit grating. The number $N$ to be factored is encoded in the number of slits of period $d$. For a fixed separation $z$ of the screen from the grating we now vary the wavelength $\lambda$ of the light. The factors of $N$ emerge for those wavelengths where all maxima of the intensity pattern are equal in height. In Appendix \ref{N-slit interferometer} we show that in this case the intensity distribution follows from the Gauss sum

\be
\mathcal{G}(\xi)\equiv\sqrt{\frac{1}{ l }} \sum_{n=0}^{N-1}\exp\left[-2\pi \rmi \left[\frac{n}{l} \xi -\frac{n^2}{2l}\right] \right]\label{eq:young}
\ee
where $\xi \equiv x/d$ is the scaled position on the screen and $l \equiv \lambda z/d^2$ is the dimensionless Talbot distance.

We emphasize that in contrast to the sum ${\cal A}_N^{(M)}$ defined by \eq{gauss2.2} the sum $\mathcal{G}$ given by \eq{eq:young} contains $N$ rather than $M+1$ terms. Moreover, N does not enter into the phase factors of $\mathcal{G}$.

In Ref.\cite{summhammer} the idea of factorization with a $N$-slit interfero\-meter \cite{clauser:1996} was translated  into  one with a single Mach-Zehnder interferometer.   Furthermore, Ref.\cite{summhammer} suggests a way to test several trial factors,  by connecting several Mach-Zehnder interferometer.

Another  proposal \cite{mack:2002:a,mack:2002:b,merkel:2006:a,merkel:2006:b} is based on wave packet dynamics in anharmonic  potentials giving rise to quadratic phase factors. Here, the  factors of an appropriately encoded number $N$ can be extracted from the autocorrelation function.  This quantity is in the form of a Gauss sum and experimentally accessible.

This approach is closely related to ideas  \cite{ivanov:2003,ivanov:2004} to use rotor like systems with a quadratic energy spectrum for quantum computing. Initially the wave packet is localized in space. At the revival time $T_{rev}$ the packet is identical to the initial one. However, it is  not the revivals but rather the fractional revivals which are interesting in the context of factorization. At times $t= T_{rev}/N$, with $N$ odd, the wave is localized at $N$ equally spaced positions. If $N$ is even, there exist only $N/2$ such positions. This phenomenon can be used to build a K-bit quantum computer \cite{ivanov:2003,ivanov:2004} with $K\approx log_2 N$.
It can also be employed for factorization \cite{harter:2001:a}. For times $t= l/N$, with $N$ odd, there exist $N$ maxima of the probability distribution if $N$ and $l$ are coprime. But if they share a common factor $p$ that is $l=p \cdot r$ and $N=p \cdot q$, we can only observe $q$ maxima.

We conclude our discussion of alternative proposals of Gauss sum factorization by briefly mentioning the idea \cite{ng:2009} of using two Josephson phase qbits which are coupled to a superconducting resonator. Here, the time evolution generates a phase shift of $\phi_k = 2\pi k^2\;N/\ell$. With the help of a quantum phase measurement, the Gauss sum can be calculated.

\subsection{Overview}

The article is organized as follows. 
In Sec.~\ref{gauss:factor} we introduce the continuous  Gauss sum and demonstrate its  potential to factor numbers using two examples. Moreover,  a remarkable scaling relation  of this type of Gauss sum allows us to use the realization of the Gauss sum for the number $N$ to obtain information on the factors of another number $N'$. We then turn in  Sec.~\ref{gauss:discrete} to an  investigation of the properties of the continuous Gauss sum  for integer arguments. This approach connects the Gauss sums central to the present article to the standard Gauss sums discussed in the mathematical literature. In this way we establish rules how to identify factors of a number using the continuous Gauss sum evaluated at integers. We dedicate in Sec.~\ref{Gauss:2} to yet another type of Gauss sum and connect it  to the continuous Gauss sum at integer values. Section~\ref{conclusion} summarizes the key results which are relevant for part II of this series of articles.

In order to keep the paper self-contained we have included several appendices. In Appendix \ref{algorithm for Gauss phase} we outline the quantum algorithm \cite{vandam1,vandam2,vandam3} for evaluating the phase of a specific class of Gauss sums.
In order to provide a comparison with the ideas proposed in the present article, we then in Appendix \ref{N-slit interferometer} briefly summarize the essential ingredients of the approach to factor numbers with   a $N$-slit interferometer \cite{dowling:1991} .
We conclude in Appendix \ref{ss:W(c+1/2)} by deriving the absolute value of a specific finite Gauss sum that is central to the discussion of revivals, and the factorization technique based on the continuous Gauss sum and  the $N$-slit interferometer.

\section{Continuous  Gau{ss} sum}
\label{gauss:factor}

In the present section we study the periodicity properties of the continuous Gauss sum 
\be
{\cal S}(\xi;A,B)\equiv\sumlim_{m=-\infty}^\infty  w_m
\exp\left[2\pi \rmi\left(\frac{m}{A}+\frac{m^2}{B}\right)\xi\right]
\label{eq:cals}
\ee
with respect to the possibility of factoring numbers. Here, $A$ and $B$ denote two real numbers and the argument $\xi$ assumes real values. The weight factors $w_m$ are centered around $m=0$ and are slowly varying as a function of $m$.

We establish three results: (i) when $B/A$ is an integer we can use the continuous Gauss sums $S$ to factor numbers. (ii) Maxima of $|{\cal S}(\xi;A,B)|$ located at an integer $\xi$ allow us to identify factors. (iii) By an appropriate scale transformation we can factor any number $N$ using the Gauss sum corresponding to another number $N'$.

In order to derive this result we first motivate these properties of the Gauss sum and then derive a new representation of this Gauss sum. This function allows us to choose the parameters appropriate for this technique. Moreover, we show that it is also possible to factor even numbers.

\subsection{A tool to factor numbers\label{sec:fac_example}} 

\begin{figure}[ht!]
\begin{center}
\includegraphics[width=0.65\columnwidth]{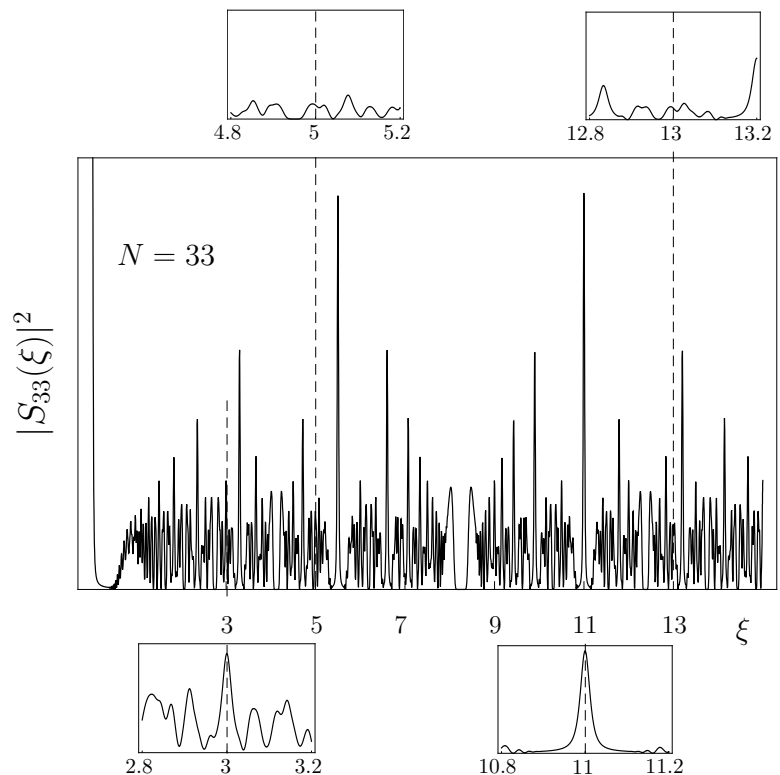}
\end{center}
\caption{
Factorization of the odd number $N=33=3 \cdot 11$ with the help of the continuous Gauss sum
$|S_{33}(\xi)|^2$ given by \eq{gauss:sum}.
The insets magnify the behavior of $|S_{33}(\xi)|^2$ in the immediate neighborhood of test factors. The pronounced maxima at the factors $3 \textrm{ and }11$ are clearly visible in the insets at the bottom. In contrast, at non-factors the signal does not show any peculiarities as exemplified at the top by $\ell=5$ and $\ell=13$.
The width of the Gaussian weight function, \eq{eq:gaussocc}, is given by $\Delta m=10$ and the summation over $m$ covers $2M+1=81$ terms.
}
\label{figure33}
\end{figure}

We start our analysis of the continuous Gauss sum by considering the special case $A=1$ and $B=N$, that is 
\be
S_N(\xi)\equiv
\sum\limits_{m=-\infty}^{\infty} w_m
\exp\left[2\pi\,\rmi\left(m+\frac{m^2}{N}\right)\xi\right].
\label{gauss:sum}
\ee
In \fig{figure33} we show $|S_N(\xi)|$ for the example $N=33=3\cdot 11$ and the Gaussian weight factors 
\be
w(m)\equiv\sqrt{\frac{1}{2\pi\Delta m^2}}\exp\left[-\frac{1}{2}\left(\frac{m}{\Delta m}\right)^2\right]\label{def:w(m)}
\ee
of width $\Delta m=10$.

Our main interest is in the behavior of $|S_N(\xi)|^2$  in the vicinity of candidate prime factors $\xi=\ell$. Indeed, the insets of \fig{figure33} at the bottom bring out most clearly  distinct maxima at values of $\xi$ corresponding to the factors $\ell=3$ and $\ell=11$. At non-factors such as $\ell=5$ or $\ell=13$ illustrated by the insets on the top, the continuous Gauss sum $S_{33}$ does not show any peculiarities. This example suggests that the continuous Gauss sum $S_N$ represents a tool for factorization.

\begin{figure}[ht!]
\begin{center}
\includegraphics[width=0.5\columnwidth]{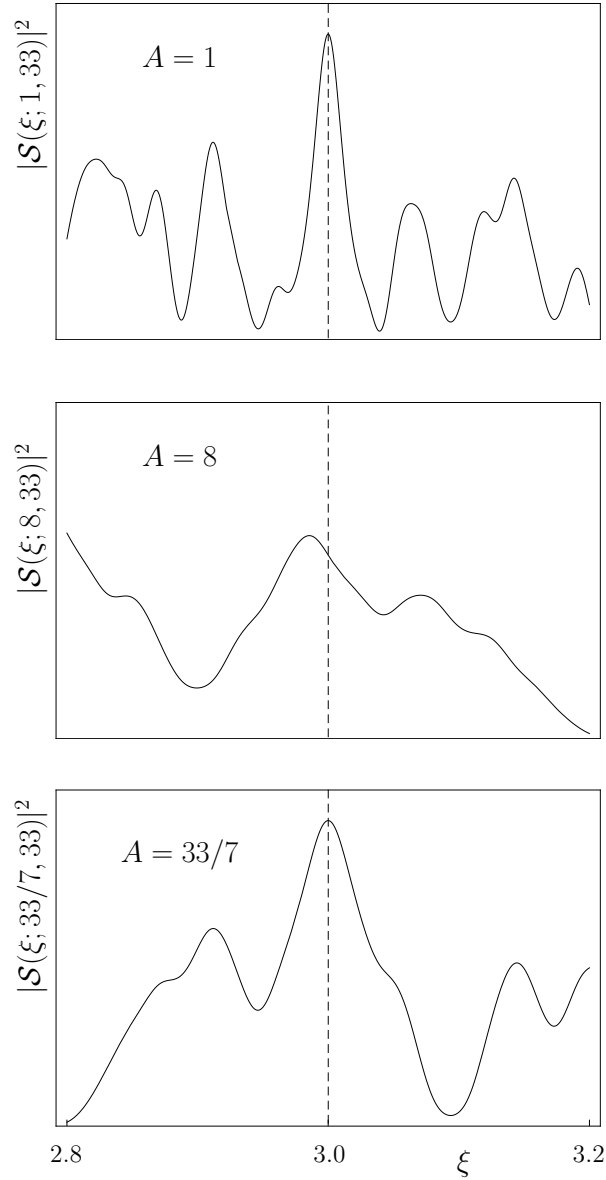}
\end{center}
\caption{Role of the parameter $A$ in the method of factoring a number with the help of the continuous Gauss sum ${\cal S}(\xi;A,B)$ defined by \eq{eq:cals}. We depict $|{\cal S}(\xi;A,33)|^2$
in the vicinity of the factor $\xi=3$ for three  different values of $A$ and note that for $A=1$ (top) and $A=33/7$ (bottom) the quantity $|{\cal S}|^2$ shows a distinct maxima at $\xi=3$. However, for $A=8$ (middle)  it does not display any peculiarities.
}
\label{Variation_A}
\end{figure}

We now turn to the role of  the parameter $A$ in the continuous Gauss sum ${\cal S}$.
For this purpose, we analyze in  \fig{Variation_A} the function $|{\cal S}(\xi;A,B)|^2$ for the  example $B=33$ and  different arguments $A=1,8$ and $33/7$ in the vicinity of the factor $\xi=3$ of $N=B=33$. For the case $A=8$ (middle) the distinct maxima at the factor $\xi=3$ of $N=33$  is not visible anymore. On the other hand, for $A=33/7$ (bottom) the absolute value $|{\cal S}(\xi;A,B)|^2$ shows the same behavior as $S_N(\xi)$. As a consequence, the continuous Gauss sum ${\cal S}(\xi;A,B)$ allows  factorization only if $B/A$ is an integer number.

\begin{figure}[ht!]
\begin{center}
\includegraphics[width=0.65\textwidth]{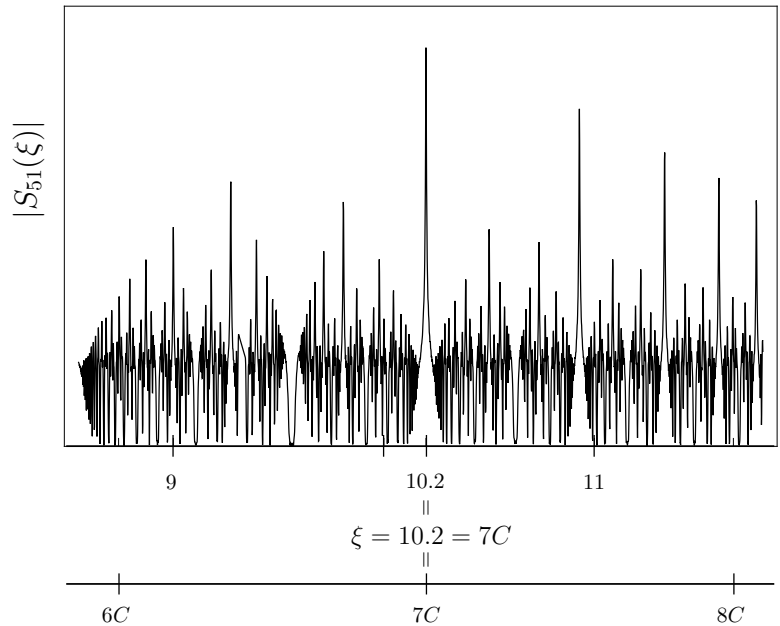}
\end{center}
\caption{Scaling property of the continuous Gauss sum illustrated by the example $|S_{51}(\xi)|$. The  dominant maximum at $\xi=10.2$  does not correspond to a factor of $N=51$ because $10.2$ is not an integer. However, by rescaling the horizontal axis, that is the variable $\xi$ with the help of a new unit $C$ we can identify this maximum as the factor of a different numbers $N'$. Indeed, the choice  $\xi=10.2\equiv7C$ with $C\equiv 51/35$ yields the  factor $7$ of $N'\equiv N/C=B/C=35$. The lower axis indicates this new scaling.
}\label{Masstab}
\end{figure}

Furthermore, \fig{figure33} displays maxima at non-integer values of $\xi$ as well. For example, we find that $|S_{51}(\xi)|$ shows a maxima at $\xi=10.2$ as indicated by \fig{Masstab}. Because $\xi=10.2$ is a non-integer, it is  thus not relevant in the context of factoring the number $N=51$. However, by rescaling the argument $\xi$ with
\be
\xi\equiv\xi'\cdot C\label{replacement}
\ee
and $C=51/35$, we find that this maxima corresponds  to the integer number $\xi'=7$. This feature is remarkable, because $\xi'=7$ is a factor of $N'\equiv 51/C=35$. As a consequence, the signal $|S_{N}(\xi)|$ does not only contain the information about the factors of $N$ but also of other numbers $N'$.

This scaling property of the Gauss sums suggests  to store a master curve $|S_N(\xi)|$  on a card small enough to be carried in the pocket which can be used to factor any other number $N'$ which is of the same order as $N$. For this reason the scaling property of the Gauss sum and the associated possibility to factor many numbers has been jokingly called \cite{stenholm}  ``pocket factorizator''. 

\begin{center}
\begin{figure}[ht]
\begin{center}
\includegraphics[width=0.85\textwidth]{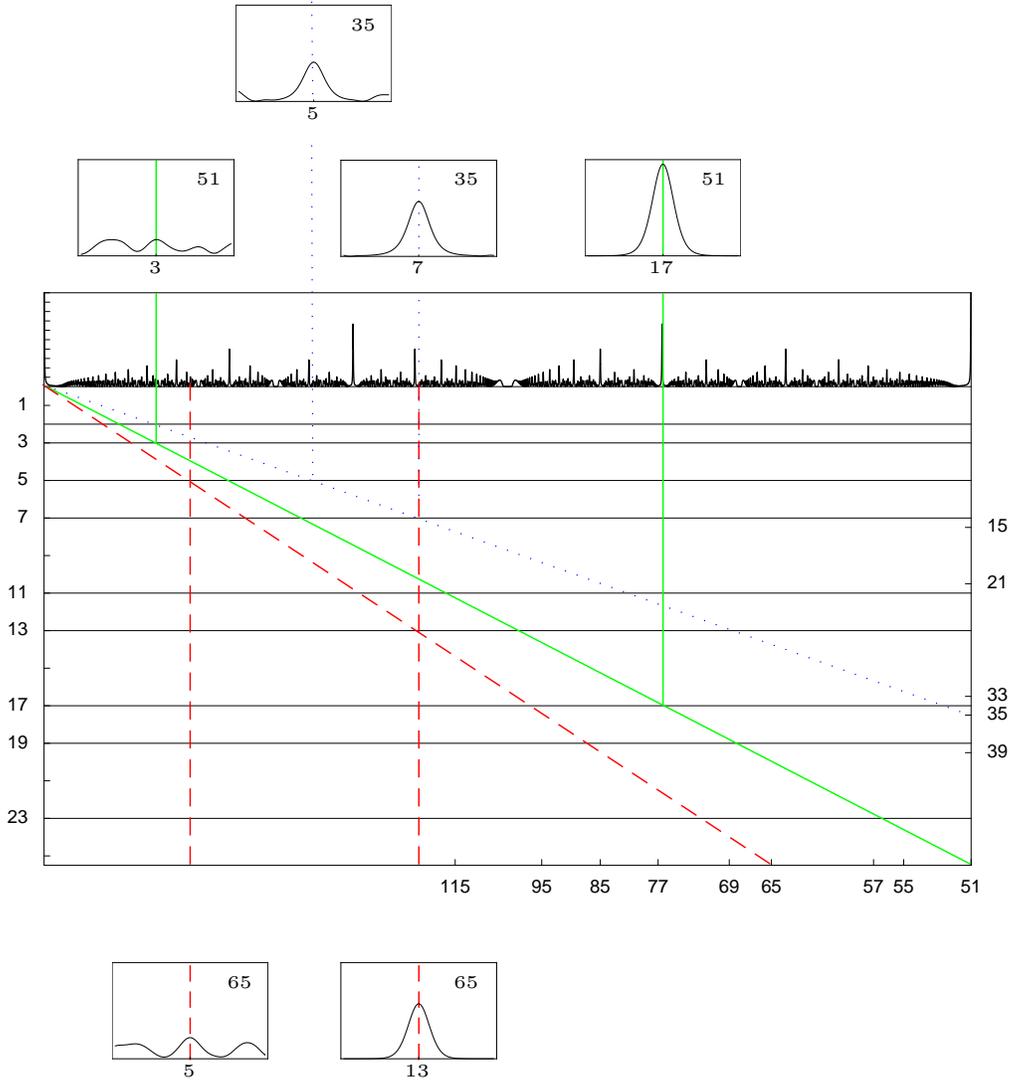}
\end{center}
\caption{The working principles of the  pocket factorizator illustrated by the factorization of the numbers $N'=35$ and $N'=65$ using the Gauss sum $S_N$ for $N=51$. 
On the right vertical axis and the horizontal bottom axis we show the numbers $N'$ to be factored in the domains $N'<N=51$ and  $N'>N=51$, respectively.
In order to factor the number $N'$ using the master signal $|S_{51}(\xi)|^2$ shown in the top row,  we need to rescale the horizontal axis according to the transformation $\xi'\equiv  (N'/N)\cdot\xi $. We achieve this task nomographically by connecting the left upper corner marking the origin of the master signal with the point $N'$ located either on the right vertical axis, or the horizontal bottom axis, by a straight line. We illustrate this procedure for the numbers $N'=35$, $N=51$ and $N'=65$ by dotted, solid, and dashed lines respectively. To  test whether the prime argument $\ell$ is a factor of $N'$ we follow the horizontal lines representing the prime numbers marked on the left vertical axis to the intersection with the tilted line. The ordinate of the intersection point defines the relevant part of the signal.
The insets show the magnified signal in the neighborhood of the prime factors of $N$ and $N'$.
A maximum of the signal at an integer identifies a factor as indicated by the insets for $N'=35$ or $N=51$ (top), and $N'=65$ (bottom).
\newline}
\label{pocketfactorizer}
\end{figure}
\end{center}

In \fig{pocketfactorizer} we illustrate the working principle of the pocket factorizator for $N' = 35$ and  $N'=65$ starting from the original signal associated with $N=51$. 
Candidate prime factors indicated on the left vertical axis are marked  by horizontal lines.
In order to find the adequate scale $\xi'$ given by \eq{replacement} for the number $N'$ we adapt the slope of the tilted line appropriate for $N'$. On the right vertical axis,   we depict numbers $N'<N$ and on the horizontal axis numbers $N<N'$.
In order to test whether the prime argument $\ell$ is a factor of $N'$ we follow the horizontal line representing $\ell$  to the intersection with the tilted line. The ordinate $\xi$  of the intersection point yields the section of the signal which has to be analyzed. When the signal displays a maximum at an integer value of $\xi'$  we have found a factor of $N'$ as indicated by magnified insets of the signal.

\subsection{A new representation}

In the preceding section we have shown for specific choices of the parameters $A$ and $B$  that the continuous Gauss sum ${\cal S}(\xi;A,B)$ allows us to factor numbers. In the following sections,  we now verify this property in a rigorous way.

For this purpose, 
we rewrite the sum, \eq{eq:cals}, in an exact way as to bring out the features of ${\cal S} ={\cal S}(\xi;A,B)$ typical for the different domains of $\xi$.
Here, we  concentrate on arguments
 \be
\xi\equiv\frac{q}{r} B+ \delta 
\label{ansatz}
\ee
that are close to a fraction $q/r$ of $B$. We use a method which has been developed in the context of fractional revivals of wave packets \cite{leichtle:PRA:1996,schleich:2001:thebook}.

When we substitute the representation \eq{ansatz} of $\xi$ into the term of the Gauss sum which is quadratic in $m$ we find
\be
{\cal S}=\sum\limits_{m=-\infty}^{\infty}w_m\exp\left[2\pi \rmi \frac{q}{r}m^2\right]\exp\left[2\pi \rmi (\frac{\xi}{A}m+\frac{\delta}{B}m^2)\right].\label{eq:S_um_B}
\ee
Next, we recall the representation 
\be
\sum\limits_{m=-\infty}^{\infty}a_m=\sum\limits_{p=0}^{r-1}\sum\limits_{n=-\infty}^{\infty}a_{p+nr}\label{eq:sumsum}
\ee
of one sum by a sum of sums which yields
\begin{eqnarray}
\fl{\cal S}=\sum\limits_{p=0}^{r-1}\sum\limits_{n=-\infty}^{\infty}w_{p+nr} \exp\left[2\pi \rmi \frac{q}{r}(p^2+2pnr+r^2)\right]\nonumber \\
\exp\left\{ 2\pi \rmi \left[ \frac{\xi}{A}(p+nr)+\frac{\delta}{B}(p+nr)^2\right] \right\},
\end{eqnarray}
or
\begin{eqnarray}
\fl {\cal S}=\sum\limits_{p=0}^{r-1}\exp\left[2\pi \rmi \frac{q}{r}p^2\right]  \sum\limits_{n=-\infty}^{\infty}w_{p+nr}  
\exp\left\{ 2\pi \rmi \left[ \frac{\xi}{A}(p+nr)+\frac{\delta}{B}(p+nr)^2\right] \right\}.
\end{eqnarray}
Here, we have made use of the identity
\be
\exp\left[2\pi \rmi s\right]=1
\ee
for integer $s$.

With the help of the Poisson summation formula
\be
\sum\limits_{n=-\infty}^{\infty}f_n = \sum\limits_{m=-\infty}^{\infty}\;\int\limits_{-\infty}^{\infty} \hspace*{-0.4 em}\rmd\nu\; f(\nu)e^{-2\pi \rmi m \nu}\label{def:poisson}
\ee
where $f(\nu)$ denotes the continuous extension of $f_n$ with $f(n)\equiv f_n$ for integer values of $n$ we arrive at
\begin{eqnarray}
\fl{\cal S}= \sum\limits_{p=0}^{r-1}\exp\left[2\pi \rmi \frac{q}{r}p^2\right] \sum\limits_{m=-\infty}^{\infty}\;\int\limits_{-\infty}^{\infty}  \hspace*{-0.4 em}\rmd\nu\; w(p+\nu r)\nonumber\\
 \exp\left\{2\pi \rmi \left[ \frac{\xi}{A}(p+\nu r)+\frac{\delta}{B}(p+\nu r)^2-m\nu\right]\right\}.
\end{eqnarray}
Here, $w(\mu)$ denotes the continuous extension of $w_m$ with $w(m)\equiv w_m$.

The substitution $\mu \equiv p+\nu r$ finally leads us to
\begin{eqnarray}
\fl{\cal S}=\sum\limits_{m=-\infty}^{\infty}\frac{1}{r} \sum\limits_{p=0}^{r-1}\exp\left[2\pi \rmi \left(\frac{q}{r}p^2+\frac{m}{r}p\right)\right]\\ \int\limits_{-\infty}^{\infty}  \hspace*{-0.4 em}\rmd\mu\; w(\mu) \exp\left[ 2\pi \rmi \left(\frac{\xi}{A}-\frac{m}{r}\right)\mu\right] \exp\left[ 2\pi \rmi \frac{\delta}{B}\mu^2\right]
\end{eqnarray}
where we have also interchanged the  summations over $m$ and $p$.

As a consequence, we can represent ${\cal S}(\xi;A,B)$ in the form
\be
{\cal S}(\xi;A,B)=\sumlim_{m=-\infty}^\infty 
{\cal W}_m^{(r)}\,{\cal I}_m^{(r)}(\xi;A,B)
\label{eq:summewmim}
\ee
with the finite Gauss sum  \cite{lang:1970}
\be
{\cal W}_m^{(r)}\equiv\frac{1}{r}\sumlim_{p=0}^{r-1}
\exp\left[2\pi \rmi\,\left(\frac{q}{r}p^2+\frac{m}{r}p\right)\right]
\label{eq:wmrgausssumme}
\ee
and the shape functions

\begin{eqnarray}
{\cal I}_m^{(r)}(\xi;A,B)\equiv
\int\limits_{-\infty}^\infty \rmd\mu\; w(\mu)\exp\left[
2\pi \rmi\left(
\frac{\xi }{A}-\frac{m}{r}\right)\mu\right] 
\times\exp\left[2\pi \rmi \frac{\delta}{B}\mu^2
\right].
\label{eq:imrofdeltat}
\end{eqnarray}

\subsection{Location and origin of maxima \label{sec:localizing_maxima}}

According to \eq{eq:summewmim} the continuous Gauss sum ${\cal S}$ in the neighborhood of  $\xi\cong q \; B/r$ consists of a sum the products ${\cal W}_m^{(r)}{\cal I}_m^{(r)}$  of the finite Gauss sum ${\cal W}_m^{(r)}$ and the shape function ${\cal I}_m^{(r)}$. Its role
stands out most clearly for the example of the continuous extension 
\be
w(\mu)\equiv\sqrt{\frac{1}{2\pi\Delta m^2}}\exp\left[-\frac{1}{2}\left(\frac{\mu}{\Delta m}\right)^2\right]
\label{eq:gaussocc}
\ee
of the Gaussian weight function $w_m$ given by \eq{def:w(m)}.

In this case we can perform the integral, \eq{eq:imrofdeltat}, and find the complex-valued Gaussian

\be
{\cal I}_m^{(r)}={\cal N}
\exp\left[-\left(\frac{m-\bar{m}}{\sigma}\right)^2(1+\rmi D\delta)\right]\label{def:I_m}
\ee
of width
\be
\sigma^2(\delta)\equiv \sigma_0^2(1+D^2\delta^2)\label{def:sigma(delta)}
\ee
and centered around
\be
\bar{m}\equiv q\frac{B}{A}+\frac{r}{A}\delta.\label{def:m_bar}
\ee
Here, we have introduced the abbreviations
\be
\sigma_0^2\equiv\sigma^2(0)\equiv \frac{r^2}{2\pi^2\Delta m^2}\label{def:sigma_0}
\ee
for the width at $\delta=0$ as well as
\be
D\equiv 4\pi  \frac{\Delta m^2}{B}\label{def_D},
\ee
together with the normalization constant
\be
{\cal N}\equiv \sqrt{\frac{1}{1-iD \delta}}. 
\ee

It is the size of the width $\sigma$ which determines how many terms contribute to the sum over $m$ in \eq{eq:summewmim}. Indeed, when $\sigma<1$, that is, for a narrow Gaussian only the $m$-value closest to $\bar{m}$ contributes while for $1<\sigma$, that is for a broad distribution shape function ${\cal I}_m^{(r)}$ of many neighboring $m$-values have to be added up in order to yield ${\cal S}$. In this case the fact that ${\cal I}_m^{(r)}$ is a complex-valued Gaussian with a quadratic phase variation as expressed by the last term in the exponential of \eq{def:I_m} becomes important.

Indeed, we recall from Appendix \ref{ss:W(c+1/2)} that the absolute value $\left|{\cal W}_m^{(r)}\right|$ of the finite Gauss sum is either constant as a function of $m$ or oscillates between a constant and zero depending on $r$ being odd or even. As a result all terms ${\cal I}_m^{(r)}$ contribute with equal weight or not at all. Although the phases of ${\cal W}_m^{(r)}$ vary rapidly \cite{berry:2001} with $m$ they cannot compensate the quadratic variation in $m$ of the phase factor governing the shape function ${\cal I}_m^{(r)}$. As a result the sum over several $m$-values leads to a destructive interference and a small value for ${\cal S}$.

\begin{figure}[ht]
\begin{center}
\includegraphics[width=0.65\columnwidth]{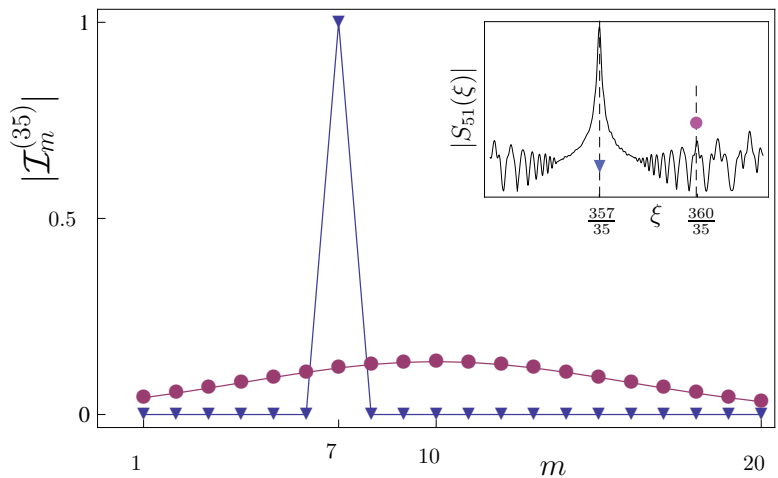}
\end{center}
\caption{
Dominant maxima in, or small values of the continuous Gauss sum $|S_N(\xi)|$ identified with the help of the sum \eq{eq:summewmim}  as originating from  a single shape function $I_m^{(r)}$, or from the  destructive interference of several of them  illustrated for $N=51$. The inset shows $|S_{51}(\xi)|$ as a function of $\xi$ which displays  a clear maximum at $\xi =357/35=(7/35)51$. Here, the deviation $\delta$  vanishes and therefore, only the term ${\cal I}^{(35)}_7$ is relevant for the Gauss sum. On the other hand, for $\xi=360/51=(7/35)51+3/35$ we find $\delta=3/35$. As a consequence, several shape functions ${\cal I}^{(35)}_m$ interfere destructively in the sum \eq{eq:summewmim} leading to a significantly reduced signal.
}
\label{fig:I_m}
\end{figure}

Figure \ref{fig:I_m} illustrates this single-maximum versus destructive interference-of-many-terms behavior for the signal $\left|S_{51}(\xi)\right|^2$ depicted in the inset.
The dominant maximum at $\xi=(7/35)51$ marked by a triangle arises solely  from the term $I_7^{(35)}$. On the other hand, for $\xi=(7/35)51+3/35$  several terms $I_m^{(35)}$ depicted by filled circles interfere destructively leading to a suppression of the signal.

Hence, the width $\sigma$ of the Gaussian shape function \eq{def:I_m} governs the value of ${\cal S}$. According to the definition \eq{def:sigma(delta)} of $\sigma$ the smallest value of $\sigma$ appears for $\delta=0$. Provided $\sigma_0\ll 1$ a single term in the sum, corresponding to  the $m$-value $m'$ closest to $\bar{m}$ will contribute. In this case the size of the shape function ${\cal I}_{m'}^{(r)}$ is governed by $\exp\left[-(m'-\bar{m})^2/\sigma_0^2\right] $. Obviously the largest signal arises when $\bar{m}$ is an integer $m_0$ since then the argument of the Gaussian  ${\cal I}_{m_0}^{(r)}$ vanishes. According to the definition \eq{def:m_bar} of $\bar{m}$ we find with $\delta=0$ for this optimal case the condition

\be
m_0=q\frac{B}{A},
\ee
which for $q=1$ yields the condition that B/A must be an integer, otherwise we cannot obtain a maximum.

We conclude by noting that due to the definition \eq{ansatz} of $\xi$ the condition $\delta=0$ for a maximum to occur translates  into the value
\be
\xi=\frac{q}{r}B,\label{eq:xi_B}
\ee
for the location. Hence,  maxima of the continuous Gauss sum ${\cal S}$ appear for integer multiples of $B/r$. However, they only emerge provided $\sigma_0\ll 1$ and B/A is an integer.

\subsection{Condition for destructive interference}

In the previous section we have identified the decisive role of the width $\sigma_0$ in allowing for dominant maxima in ${\cal S}$. In particular, we have established the necessary condition $\sigma_0\ll1$ for the occurrence of maxima. We now derive a criterion for the destructive interference of many shape functions ${\cal I}_m^{(r)}$ in the sum \eq{eq:summewmim}. 

This property arises from the condition
\be
1\ll\sigma(\delta)=\sigma_0(1+D^2\delta^2)^{1/2}
\ee
on the width which implies

\be
\left[\frac{1}{\sigma_0^2}-1\right]\ll D^2\delta^2\label{eq:sigma-1}.
\ee

When we recall the constraint $\sigma_0^2\ll 1$ necessary for the occurrence of a maximum which implies $1<<1/\sigma_0^2$ we can neglect the term unity in the square and require the condition

\be
\frac{1}{\sigma_0^2}\ll D^2\delta^2,
\ee
which is even more general than the inequality \eq{eq:sigma-1}.

With the help of the definitions \eq{def:sigma_0} and \eq{def_D} of $\sigma_0$ and $D$ we find
\be
\frac{1}{r^2}\ll 8\Delta m^2\left(\frac{\delta}{B}\right)^2 .\label{eq:m(delta,B)}
\ee
As a consequence, the width $\Delta m^2$ depends on the minimal value of $\delta$ which we want to discriminate from zero. For factoring the number $N$ we estimate the continuous Gauss sum at arguments
\be
\xi=\ell C
\ee
with $C=B/N$. As a consequence, the parameter $\delta$ from \eq{ansatz} is given by
\be
\delta = \xi-\frac{q}{r}B= \ell \frac{B}{N} -\frac{q}{r}B=\frac{\ell \cdot r-qN}{Nr}B
\ee
and therefore, $|\delta|$ is larger than $B/(Nr)$ if it is non-vanishing. As a result, the condition of \eq{eq:m(delta,B)} reduces to
\begin{equation}
1\ll8\frac{\Delta m^2}{N^2}\label{bed_m2},
\end{equation}
which is independent of $r$. 

In summary, if $\Delta m$ is larger than $r$ and $N$, and $B/A$ is integer, we can distinguish between $\delta=0$ and $\delta\neq 0$. In this case, we see peaks if and only if $\delta=0$ which will help us to factor the number $N$ as demonstrated in the following section.

\subsection{Factorization  }

We now show, that the continuous Gauss sum ${\cal S}(\xi;A,B)$ given by \eq{eq:cals} and represented for a Gaussian weight function $w_m$ defined by \eq{def:w(m)} by a sum  of complex-valued Gaussians given by  \eq{eq:summewmim} offers a tool to factor numbers. Here, we distinguish between odd and even numbers to be factored. Needless to say, the last case is not of practical interest since we can always extract powers of 2 from an even number. Nevertheless, it is interesting from a principle point of view. In particular, it brings out the crucial role of the finite Gauss sum ${\cal W}_m^{(r)}$ in our factorization scheme.

\subsubsection{Odd numbers}

In the examples discussed in Sec. \ref{sec:fac_example} we have found factors by searching for maxima at arguments $\xi$  which are integer multiples of $C$, that is $\xi=\ell C$. In Sec.\ref{sec:localizing_maxima} we have shown that maxima correspond to arguments $\xi = (q/r)B$. As a consequence, the identity \eq{ansatz} transforms for maxima at $\xi=\ell C$ into
\be
\ell \cdot C = \frac{q}{r}B.
\ee
When we define the number $N$ to be factored by the  units $C$ and $B$ of our system, that is  $N \equiv B/C$, this equation turns into
\be
\ell = \frac{q}{r}N.
\ee
Here, we have to choose the two coprime integers $q$ and $r$ such that $\ell$ is an integer. The condition is only met if $r$ corresponds to a factor of $N$ and $r$ must be odd for odd numbers $N$.
 
As a result we have derived the criterion for the factorization of odd $N$ suggested in Figs.\ref{figure33}-\ref{pocketfactorizer}: If the Gauss sum ${\cal S}$ exhibits a maximum at the integer argument $\xi=\ell$ then $\ell$ corresponds to a prime factor, or a multiple of a factor of $N$.

\subsubsection{Even numbers}
\label{evenN}

So far we have utilized only the shape function ${\cal  I}_m^{(r)}$ to factor a number. Moreover, the technique is limited to odd numbers. We now show that the finite Gauss sum ${\cal W}_m^{(r)}$ defined by \eq{eq:wmrgausssumme} together with ${\cal I}_m^{(r)}$ yields information on the factors of $N$ when $N$ is even. 

Even though the condition $\delta = 0$ would allow for a maximum at the argument $\ell$, we  may still find a vanishing value of the continuous Gauss sum ${\cal S}$. This behavior originates from the weights ${\cal W}^{(r)}_m$ which depend critically on the classification of $r$ and $q$ according to
\be
|{\cal W}_m^{(r)}|=\left\{
\begin{array}{rlll}
\sqrt{1/r}&\textrm{ for }r\textrm{ odd}\\[3mm]
\sqrt{2/r}&\textrm{ for }r\textrm{ even, }& rq/2\textrm{ even and }&m \textrm{ even}\\
0         &\textrm{ for }r\textrm{ even, }& rq/2\textrm{ even and }&m \textrm{ odd}\\[3mm]
0         &\textrm{ for }r\textrm{ even, }& rq/2\textrm{ odd and }& m \textrm{ even}\\
\sqrt{2/r}&\textrm{ for }r\textrm{ even, }& rq/2\textrm{ odd and }& m \textrm{ odd}
\end{array}
\right. 
\label{wmr:result}
\ee
derived in  Appendix \ref{ss:W(c+1/2)}.

Indeed, the weights ${\cal W}_m^{(r)}$ in the representation \eq{eq:summewmim} of ${\cal S}$ vanish for specific combinations of $r$ and $q$ thus leading to a suppression of the signal at certain integer arguments. For even $r$ and even $rq/2$ the finite Gauss sum ${\cal W}_m^{(r)}$ vanishes for odd values of the summation index $m$, whereas for even $r$ and odd $rq/2$ we find that ${\cal W}_m^{(r)}$ vanishes only for even values of  $m$.

\begin{figure}[ht]
\begin{center}
\includegraphics[width=0.6\columnwidth]{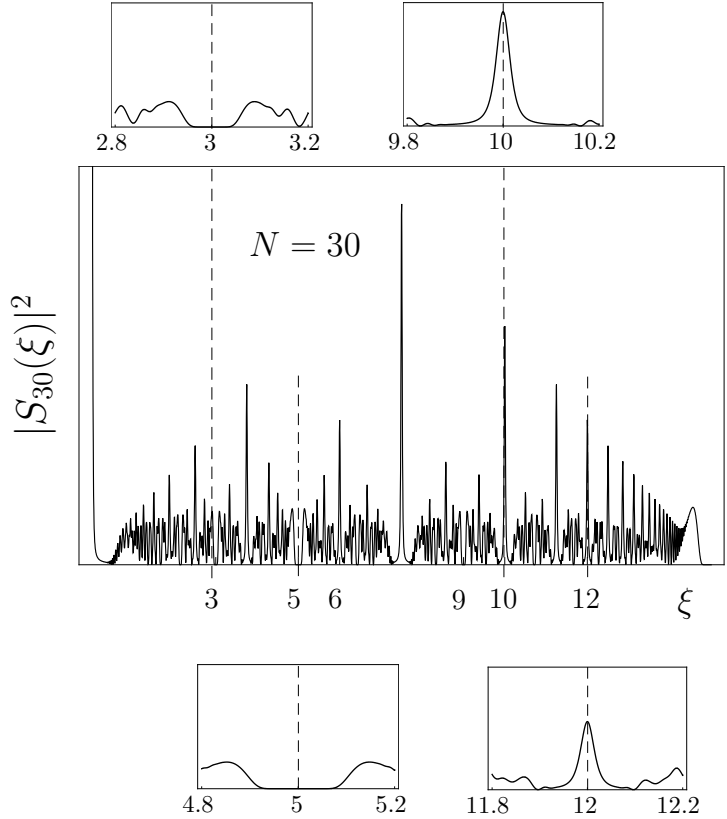}
\end{center}
\caption{
Factorization of the even number $N=30 = 2 \cdot 3 \cdot 5$ with the help of the Gauss sum $|S_{30}(\xi)|^2$ given by\eq{gauss:sum}.
The insets magnify the behavior of $|S_{30}(\xi)|^2$ in the immediate neighborhood of an integer.  Factors such as $3$ and $5$ display zeros as exemplified by the insets on the left. However, integers containing products of factors lead to maxima in $| S_{30}(\xi)|^2$ as shown by the insets on the right for $10=2 \cdot 5$ and $12 = 2^2 \cdot 3$.The width of the Gaussian weight function, \eq{eq:gaussocc}, is given by $\Delta m=8$.
}
\label{figure30}
\end{figure}

Thus for even $N$ both, maxima {\it and} zeros of $|{\cal S}(\xi;A,B)|$ at integer arguments $\xi=\ell$ contain information about the factors of $N$ as demonstrated in \fig{figure30} for the example $N=30=2\cdot 3\cdot 5$. Here, we  find a vanishing signal for $\ell=3$ and $\ell=5$ as indicated by the left insets, whereas the signal shows pronounced maxima at $\ell=10=2 \cdot 5$ and $\ell=12=3 \cdot 4$.
Obviously, these features are related to the  factors $2,3$ and $5$ of $N=30$.


\section{Discrete Gauss sum}
\label{gauss:discrete}

So far we have analyzed the continuous Gauss sum ${\cal S}$ and the special case $S_N$ in its dependence on the argument $\xi$, which assumes real numbers. We now restrict $S_N$ to integer arguments \mbox{$\xi \equiv \ell$} and recall the identity  $\textrm{exp}\left( 2\pi i m \ell \right)=1$. As a consequence,  the continuous Gauss sum \eq{gauss:sum} reduces to 
\be
S_N(\ell)=
\sum\limits_{m=-\infty}^{\infty} w_m
\exp\left[2\pi\,\rmi\frac{m^2}{N}\ell\right].
\label{gauss:at:int:arg}
\ee
In the present section we show that the restriction to integer arguments allows us to derive analytical expressions for $S_N(\ell)$ in terms of  the standard Gauss sum 
\be
G(a,b)\equiv\sum\limits_{m=0}^{b-1} \exp\left(2\pi \rmi\, m^2\frac{a}{b}\right),
\label{standard}
\ee
where  $a$ and $b$ denote integers. Its periodicity properties  provide us with rules how to factor numbers based on $S_N(\ell)$.

\subsection{New representation}
For this purpose we first cast $S_N(\ell)$ into a new form which can be approximated in the limit of a broad weight function $w_m$  by the standard Gauss sum $G$. Indeed, the representation  \eq{eq:sumsum} of one sum by a sum of sums  yields the expression
\be
S_N(\ell)=
\sum\limits_{m=0}^{N-1}\exp\left[2\pi \rmi\,m^2 \frac{\ell}{N}\right]
\;
\sum\limits_{n=-\infty}^\infty w_{m+n N}.
\ee
We evaluate the  sum over $n$ with the help of the Poisson summation formula \eq{def:poisson} and find
\be
S_N(\ell)=\frac{1}{N}\sum\limits_{\nu=-\infty}^\infty \sum\limits_{m=0}^{N-1} 
\exp\left[2\pi \rmi \left(m^2\frac{\ell}{N}+\frac{m\,\nu}{N}\right)\right]
\tilde{w}\left(\frac{\nu}{N}\right)
\label{peg:FT:weight}
\ee
where we have introduced the Fourier transform 
\be
\tilde{w}\left(x\right)\equiv \int\limits_{-\infty}^\infty \rmd\mu\; w(\mu)\, \exp\left(-2\pi \rmi\, x\mu \right)
\ee
of the continuous extension $w(\mu)$ of the weight factors.

So far the calculation is exact. However, we now make an approximation which  connects the sum $S_N(\ell)$ given by \eq{gauss:at:int:arg} to the standard Gauss sum $G$ defined by \eq{standard}. For this purpose we recall that
according to the Fourier theorem the product of the widths $\Delta m$ and $\Delta x$ of the weight function $w_m$ and its associated Fourier transform $\tilde{w}$ is constant.
Since $w_m$ is very broad the  distribution $\tilde{w}\left(x\right)$ must be very narrow. Together with the fact that the argument of $\tilde{w}$ is $\nu/N$ with $1\ll N$ the Fourier theorem allows us to restrict  the sum over $\nu$ to the term $\nu=0$ only and we arrive\footnote{This reduction of the sum to $\nu=0$ stands out most clearly for the case of the Gaussian weight function given by (\ref{eq:gaussocc}). }at
\be
S_N(\ell)
\cong \sum\limits_{m=0}^{N-1} \exp\left[2\pi\rm i \,m^2\frac{\ell}{N}\right] \frac{1}{N} \int \limits^{\infty}_{-\infty} \hspace{-0.5em}\rmd\mu \,w(\mu).
\label{peg:truncated}
\ee
Due to the normalization, the integral over $w(\mu)$ is equal to unity and we obtain the approximation
\be
S_N(\ell)\cong \frac{1}{N} \,G(\ell,N)\label{eq_S=G}
\ee
for the discrete Gauss sum $S_N(\ell)$   in terms of  the standard Gauss sum $G$.


\subsection{Analytical expressions\label{sec:ana_expressions}}

We now use the well-known results \cite{lang:1970,number_theory,davenport:1980,schleich:2005:primes} for the standard Gauss sum $G$ to approximate $ S_N(\ell)$. Throughout the section, we assume that $N=p\cdot r$ contains the two integer factors $p$ and $r$.
Moreover, we distinguish two cases of the integer argument $\ell$.

\subsubsection{ No common factor between $\ell$ and $N$}
In this case we can take advantage of the relation \cite{schleich:2005:primes} 
\begin{equation}
G(a,b)=\left(\frac{a}{b}\right)G(1,b)
\label{eq:gauss:rel}
\end{equation}
connecting the standard Gauss sum $G(a,b)$ of the arguments $a$ and $b$ with the elementary Gauss sum $G(1,b)$ of the arguments $a=1$ and $b$ through the Legendre symbol
\begin{equation}
\left(\frac{a}{b}\right)\equiv\left\{\begin{array}{cl}
+1&{\rm if\; there\; is\; an\;}x\; {\rm with }\;b\; \textrm{being a divisor of}\;(a-x^2)\\
-1&{\rm if\; there\; is\; no\; such\; }x\\
0&{\rm if }\;b\; \textrm{ is a divisor of}\;a
\end{array}\right. .
\label{eq:legendre}
\end{equation}
When we recall \cite{schleich:2005:primes} the expression 
\begin{equation}
G(1,b)=\left\{
\begin{array}{ccl}
(1+i)\sqrt{b}&{\rm for}\;&b\in{\cal M}_0\\
\sqrt{b}&{\rm for}\;&b\in{\cal M}_1\\
0&{\rm for}\;&b\in{\cal M}_2\\
i\sqrt{b}&{\rm for}\;&b\in{\cal M}_3
\end{array}\right. ,
\label{eq:G1q}
\end{equation}
for the elementary Gauss sum $G(1,b)$ with the sets 
\be
{\cal M}_k \equiv \{r \mid r=4s+k \quad {\rm and}\quad k=0,\dots, 3\}
\ee
consisting of integers $r=4s+k$  we obtain the  approximation 
\be 
|S_N(\ell)|^2\cong \frac{1}{N}
\left\{
\begin{array}{lcl}
2 &\textrm{for}  &N\in {\cal M}_0\\
1 &\textrm{for}  &N\in {\cal M}_1,\, {\cal M}_3 \\
0 &\textrm{for}  &N\in {\cal M}_2
\end{array}
\right. 
\label{eq:modulus2:N}
\ee
for the absolute value squared of the discrete  Gauss sum $S_N(\ell)$,   provided  $\ell$ and $N$ do not share a factor.

\subsubsection{ A common factor between $\ell$ and $N$}

Next, we turn to the case where the argument $\ell$ is an integer multiple $k$ of one of the factors of $N$, that is $\ell=k\, p$.
Now, we find from \eq{eq_S=G} the identity  
\be
S_N(\ell)=S_N(kp)\cong \frac{1}{N}G(kp,rp)
\ee
which with the help of the factorization relation 
\be
G(a,b)=p\, G\left(\frac{a}{p},\frac{b}{p}\right)\label{eq:G2q}
\ee
of the standard Gauss sum reduces to
\be
S_N(k\, p)
=\frac{p}{N}G(k,r)=\frac{p}{N} \left(\frac{k}{r} \right) \,G(1,r).
\ee
In the last step we have also made use of the connection formula \eq{eq:gauss:rel}. 

The explicit expression \eq{eq:G1q} for $G(1,b)$ finally yields the approximation
\be
|S_N(k\, p)|^2\cong\frac{p}{N}
\left\{
\begin{array}{lcl}
2 &\textrm{for} &r \in {\cal M}_0\\
1 &\textrm{for} &r \in {\cal M}_1,\,{\cal M}_3\\
0 &\textrm{for} &r \in {\cal M}_2
\end{array}
\right.
\label{eq:modulus:rp}
\ee
of $|S_N|^2$ at integer multiples of the factor $p$.

\subsection{Factorization}
A comparison between the explicit expressions \eq{eq:modulus2:N} and\eq{eq:modulus:rp} for $| S_N(\ell)|^2$ indicates a method to factor numbers. In order to illustrate this technique we first assume $N$ to be odd.

In this case $N$ is either an element of ${\cal M}_1$ or ${\cal M}_3$ and we find according to  \eq{eq:modulus2:N}  that $|S_N(\ell)|^2$ is given  by $1/N$ provided the argument $\ell$ and the number $N$ do not share a common factor. Since the factors must be both odd, the value of $|S_N|^2$ for $\ell$ being a multiple of a factor $p$ reads $p/N$ and is enhanced by $p$. Moreover, the values $|S_N(p)|^2$ at the factors of $N$ form a line connecting the points $(\ell=1,|S_N(1)|^2=1/N)$ and $(\ell=N,|S_N(N)|^2=1)$. Obviously for prime numbers, there are no points on this line except the start and the end point.

\begin{figure}
\begin{center}
\includegraphics[width=0.95\textwidth]{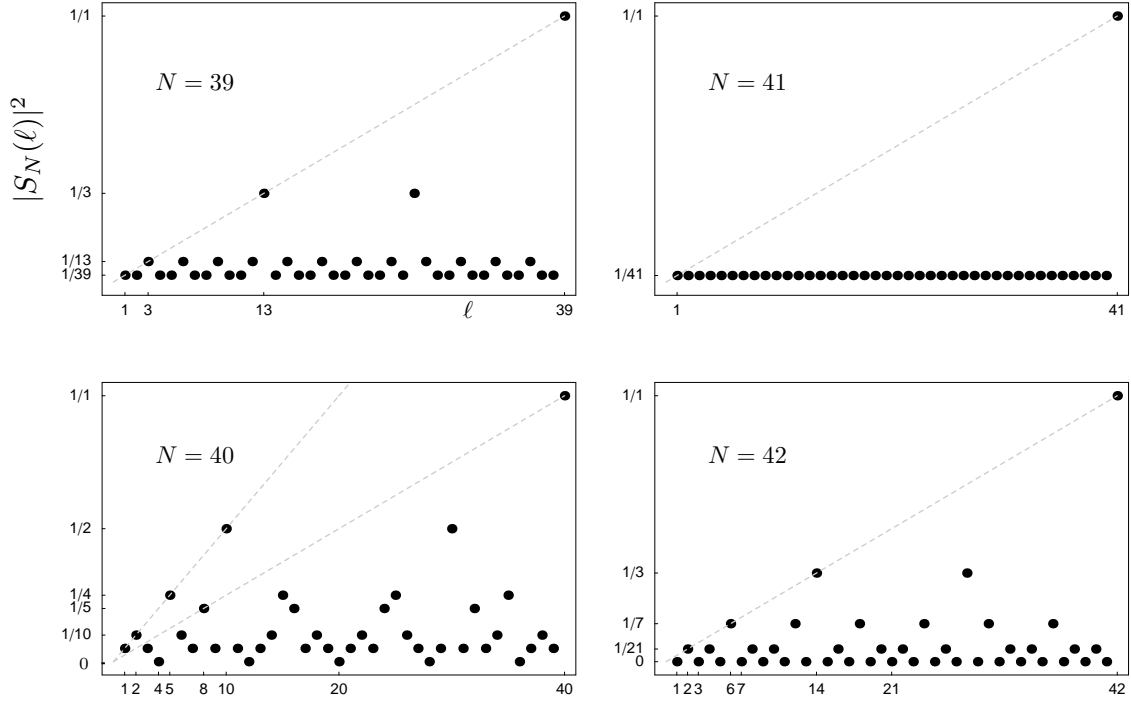}
\end{center}
\caption{Factorization of $N=39,40,41$ and $42$ using  the signal $|S_N(\ell)|^2$ corresponding to the discrete  Gauss sum $S_N(\ell)$ given by \eq{gauss:at:int:arg}. 
For $N\in {\cal M}_0$ such as $N=40$  data points corresponding to factors of $N$ arrange themselves on two straight lines through the origin. For $N\in {\cal M}_2$ exemplified by $N=42$ and odd values of $N$ such $N=39$ the factors of $N$ lie on a single straight line. 
For completeness we  also depict the case of the prime number  $N=41$ with no points on this line except $\ell=1$ and $\ell=41$. The width of the Gaussian wave function \eq{eq:gaussocc} is given by $\Delta m =10$ and the summation over $m$ covers $2M+1=81$ terms.
}
\label{39to42}
\end{figure}

These features are confirmed in  \fig{39to42} where we depict  $|S_N(\ell)|^2$ defined by \eq{gauss:at:int:arg} for the odd number $N=39=3\cdot 13=9\cdot 4+3$ which is an element of ${\cal M}_3$. No factors appear for the  prime number $N=41$. 

The situation is slightly more complicated when $N$ is even. Here, the value of $|S_N(\ell)|^2$ at non-factors is either zero if $N$ is member of ${\cal M}_2$, or $2/N$ if $N$ is member of ${\cal M}_0$. However, at multiples of the factor $p$ the value of $|S_N(kp)|^2$ can either be $2p/N$, $p/N$ or zero depending on the other factor $r$ either being a member of ${\cal M}_0$, ${\cal M}_1$ and ${\cal M}_3$, or ${\cal M}_2$, respectively. In this case, there can even be two lines of $|S_N|^2$ at factors. 

This feature stands out clearly in the example of 
$N=40=5\cdot 2^3=10 \cdot 4$ which belongs to the set ${\cal M}_0$. As a result, at non-factors we find the values $2/40=1/20$ as  shown in \fig{39to42}. At the  factor $p=5$ the remaining factor  $r=40/5=8$  is an element of ${\cal M}_0$ and the corresponding value of $|S_N|^2$ is $2\cdot 5/40=1/4$. Moreover, for the factor $p=8$ the remaining factor $r=40/8=5$ belongs to ${\cal M}_1$ and therefore yields the value $8/40=1/5$. The factor $p=20$ leads us to the remaining factor $r=40/20=2$ which belongs to ${\cal M}_2$ and creates a vanishing signal.

We conclude this discussion of factoring numbers using the discrete Gauss sum $S_N(\ell)$ by using the example of
the number $N=42=2\cdot 3 \cdot 7=10 \cdot 4+2$ which is an element of ${\cal M}_2$. As a result all non-factors have a vanishing signal. The factors   belong to the classes ${\cal M}_2$, ${\cal M}_1$ or ${\cal M}_3$. Since the class ${\cal M}_0$ does not appear, the values of $|S_N(\ell)|^2$ at factors form only a single rather than two lines.

\section{Reciprocate Gauss sum}
\label{Gauss:2}

In Secs.\ref{gauss:factor} and \ref{gauss:discrete} we have analyzed the potential of two types of Gauss sums for the factorization of numbers. Both sums share the property that the number $N$ to be factored and the variable, either in its continuous version $\xi$, or as the discrete test factor $\ell$  appear as the ratios $\xi/N$ or $\ell/N$. In the present section we investigate yet another type of Gauss sum which is even more attractive in the context of factori\-zation. Here, the roles of the variable  and the number $N$ to be factored are interchanged and the sum
\be
{\cal A}_N^{(M)}(\ell)\equiv\frac{1}{M+1}\sum\limits_{m=0}^M \exp\left(-2\pi \rmi\, m^2 \frac{N}{\ell}
\right)
\label{gauss2}
\ee
consisting of $M+1$ terms emerges. Due to the appearance of  the reciprocal of $\ell/N$ we call  ${\cal A}_N^{(M)}(\ell)$  the reciprocate Gauss sum.

We emphasize that ${\cal A}_N^{(M)}(\ell)$ has been realized in a series of experiments \cite{mehring:2007,mahesh:2007,gilowski:2008,bigourd:2008,weber:2008,peng:2008,tamma:2009,tamma:2009:b,sadgrove:2008,sadgrove:2009} and used to factor numbers as summarized in Sec.\ref{gauss:sums:physics}. However,  the application of the extension of ${\cal A}_N^{(M)}(\ell)$ to a continuous variable as a tool  for factorization is more complicated and has been analyzed in  Ref.\cite{woelk:2009}. For this reason we concentrate in this section on integer arguments $\ell$.

The sums $S_N(\ell)$ and ${\cal A}_N^{(M)}(\ell)$ are closely connected with each other. We now show that the Gauss reciprocity relation \cite{onishi:2003,hannay:berry:1980} allows us to express ${\cal A}_N^{(\ell-1)}(\ell)$ in terms of the standard Gauss sum $G$. With the help of the exact analytical expressions for $G$ we then obtain closed-form expressions for $\left|{\cal A}_N^{(\ell-1)}(\ell)\right|$ and   establish rules how to factor numbers.

\subsection{Gauss reciprocity relation}

In this section we consider the Gauss sum ${\cal A}_N^{(M)}(\ell)$ for the special choice  $M\equiv\ell-1$ for the truncation parameter $M$, that is
 we discuss the properties of the complete reciprocate Gauss sum
\be
{\cal A}_N^{(\ell-1)}(\ell)\equiv\frac{1}{\ell}
\sum_{m=0}^{\ell-1}\exp\left[-2\pi \rmi\, m^2 \frac{N}{\ell}\right].
\label{gauss:ell}
\ee
Here, for each argument $\ell$ the summation  covers $\ell$ terms.

The Gauss reciprocity relation \cite{hannay:berry:1980,onishi:2003}
\be
\sum\limits_{m=0}^{a-1}\exp\left[2 \rmi\,\pi\,m^2 \frac{b}{a} \right]=
\sqrt{\frac{ai}{2b}}
\sum\limits_{m=0}^{2b-1}\exp\left[-2 \rmi\,\pi\,m^2 \frac{a}{4b} \right]
\ee
is essential in building the connection between the complete reciprocate Gauss sum of \eq{gauss:ell}, and the standard Gauss sums $G(\ell,N)$.

When we identify $a=-\ell$ and $b=N$ we find the alternative representation
\be
{\cal A}^{(\ell-1)}_N(\ell)=\frac{1}{\ell}
\sqrt{\frac{-\rmi\ell}{2 N}}
\sum_{m=0}^{2N-1}\exp\left[2\pi \rmi\, m^2 \frac{\ell}{4N}\right].\label{eq:A_umschr_G}
\ee
At this point it is useful to take advantage of the relation
\be
\sum\limits_{k=2N}^{4N-1}\exp\left[2\pi \rmi k^2\frac{\ell}{4N} \right]=  
\sum\limits_{k=0}^{2N-1}\exp\left[2\pi \rmi m^2\frac{\ell}{4N} \right]
\ee
which follows when we introduce the summation index $m\equiv k-2N$ together with the identity $\exp(2\pi i s\ell)=1$ for integers $s$. As a result we find
\be
\sum\limits_{k=0}^{2N-1}\exp\left[2\pi \rmi m^2\frac{\ell}{N} \right] = \frac{1}{2}\sum\limits_{k=0}^{4N-1}\exp\left[2\pi \rmi k^2\frac{\ell}{N} \right]
\ee
and \eq{eq:A_umschr_G} reduces to
\be
{\cal A}_N^{(\ell-1)}(\ell)=\frac{1}{2\ell}
\sqrt{\frac{-\rmi\ell}{2 N}}
\sum_{m=0}^{4N-1}
\exp\left[2\pi \rmi\, m^2 \frac{\ell}{4N}\right]
\ee
or
\be
{\cal A}_N^{(\ell-1)}(\ell)= \frac{\textrm{e}^{-\rmi \pi/4}}{2}\,\frac{1}{\sqrt{2\ell N}}\,  G(\ell,4N).
\label{eq:A=G}
\ee
A comparison with \eq{eq_S=G} finally establishes the connection
\be
{\cal A}_N^{(\ell-1)}(\ell)\cong \rme^{-\rmi\pi/4}\sqrt{\frac{2N}{\ell}}S_{4N}(\ell)
\ee
between the complete reciprocate Gauss sum ${\cal A}_N^{(\ell-1)}(\ell)$ and the discrete Gauss sum $S_N(\ell)$ discussed in the preceding section.

\subsection{Analytical expressions}

The  connection \eq{eq:A=G} between the complete reciprocate  sum ${\cal A}^{(\ell-1)}_N(\ell)$ and the standard Gauss sum $G(\ell,4N)$ allows us to draw on the results of \eq{eq:G1q} and \eq{eq:G2q} to obtain explicit expressions for $|{\cal A}^{(\ell-1)}_N(\ell)|$.
Throughout this section we assume for the sake of simplicity that the number $N$ to be factored is odd.
In complete analogy to Sec.\ref{sec:ana_expressions} we distinguish two cases for the argument $\ell$.

\subsubsection{No common factor between $\ell$ and $N$}
With the help of the factorization relation \eq{eq:G2q} of $G$ we find 
\begin{equation}
G(\ell,4N)
=\left\{
\begin{array}{ccl}
4\,G(\frac{\ell}{4},N)&{\rm for}\;&\ell\in{\cal M}_0\\
G(\ell,4N)&{\rm for}\;&\ell\in{\cal M}_1,{\cal M}_3\\
2 \,G(\frac{\ell}{2},2N)&{\rm for}\;&\ell\in{\cal M}_2
\end{array}\right.
\end{equation}
which with the connection formula for $G(a,b)$ and $G(1,b)$,\eq{eq:gauss:rel}, together with the explicit result \eq{eq:G1q} for $G(1,b)$ leads us to
\begin{equation}
|G(\ell,4N)|
=\left\{
\begin{array}{ccl}
4\sqrt{N}&{\rm for}\;&\ell\in{\cal M}_0\\
2\sqrt{N}&{\rm for}\;&\ell\in{\cal M}_1,{\cal M}_3\\
0&{\rm for}\;&\ell\in{\cal M}_2
\end{array}\right. .
\end{equation}
Therefore, the reciprocate Gauss sum takes on the value
\begin{equation}
|{\cal A}_N^{(\ell-1)}(\ell)|
=\left\{
\begin{array}{ccl}
\sqrt{\frac{2}{l}}&{\rm for}\;&\ell\in{\cal M}_0\\
\sqrt{\frac{1}{l}}&{\rm for}\;&\ell\in{\cal M}_1,{\cal M}_3\\
0&{\rm for}\;&\ell\in{\cal M}_2
\end{array}\right.\label{A_nofactor}
\end{equation}
where we have used \eq{eq:A=G}.

\subsubsection{A common factor between $\ell$ and $N$}
Here we have $\ell=ks$ and $N=rs$, which leads us  with the help of the factorization relation \eq{eq:G2q} to
\be
G(ks,4rs)=sG(k,4r).
\ee
Furthermore, we have to analyze if $k$ and four share a common factor, which leads us to
\begin{equation}
G(ks,4rs)=\left\{
\begin{array}{ccl}4s \,G(\frac{k}{4},r)&{\rm for}\;&k\in{\cal M}_0\\
s \,G(k,4r)&{\rm for}\;&k\in{\cal M}_1,{\cal M}_3\\
2s \,G(\frac{k}{2},2r)&{\rm for}\;&k\in{\cal M}_2
\end{array}\right..
\end{equation}
In the last step, we have again made use of the factorization relation  \eq{eq:G2q}.

In the case $k \in{\cal M}_2$ it can be shown that $2r$ for each odd number $r$ belongs to the class ${\cal M}_2$. Therefore, we find
\begin{equation}
|G(ks,4rs)|=\left\{
\begin{array}{ccl}
4s\sqrt{r}&{\rm for}\;&k\in{\cal M}_0\\
2s\sqrt{2r}&{\rm for}\;&k\in{\cal M}_1,{\cal M}_3\\
0&{\rm for}\;&k\in{\cal M}_2
\end{array}\right.
\end{equation}
where we have  used \eq{eq:gauss:rel} and \eq{eq:G1q}.
Thus we find with \eq{eq:A=G} the expression
\begin{equation}
|{\cal A}_N^{(\ell-1)}(\ell)|
=\left\{
\begin{array}{ccl}
\sqrt{\frac{2}{k}}&{\rm for}\;&k\in{\cal M}_0\\
\frac{1}{\sqrt{k}}&{\rm for}\;&k\in{\cal M}_1,{\cal M}_3\\
0&{\rm for}\;&k\in{\cal M}_2
\end{array}\right. \label{eq:A_factor}
\end{equation}
for the absolute value of the Gauss sum.


\subsection{Factorization}

We now take advantage of the results obtained in the preceding section to factor an odd number $N$ with the help of the signal $|{\cal A}_N^{(\ell-1)}(\ell)|$.
For arguments $\ell$ corresponding to factors of $N$ we find the maximal value $|{\cal A}_N^{(\ell-1)}(\ell)|\,=\,1$.
However, if $\ell$ and $N$ share a common factor $s$, that is $\ell=ks$ and $N=rs$, we obtain $|{\cal A}_N^{(\ell-1)}|=\sqrt{2/k},\, \sqrt{1/k}$ or $0$.
Provided $\ell$ and $N$ have no factor in common \eq{A_nofactor} predicts
$|{\cal A}_N^{(\ell-1)}|=\sqrt{2/\ell},\,\sqrt{1/\ell}$ or $0$.

\begin{figure}
\centering

\begin{center}
\includegraphics[width=0.6\textwidth]{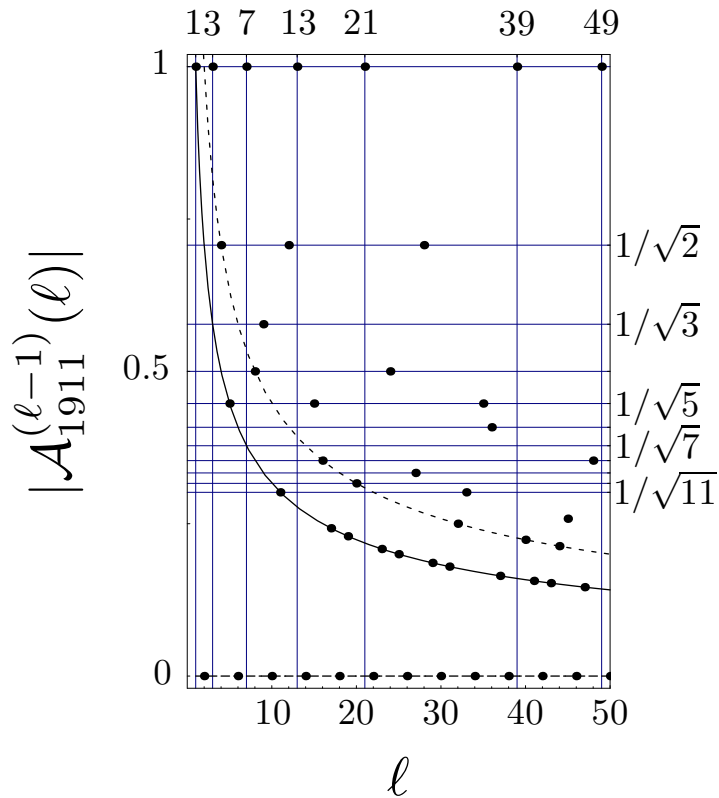}
\end{center}
\caption{Factorization of $N=1911=3\cdot 7^2 \cdot 13$ with the help of the signal $|{\cal A}_{1911}^{(\ell-1)}(\ell)|$ originating from the complete reciprocate Gauss sum defined by \eq{gauss:ell} and given explicitly by \eq{A_nofactor} and\eq{eq:A_factor}.
The positions of the factors are indicated by vertical lines. For factors the signal displays a maximum with $|{\cal A}_{N}^{(\ell-1)}(\ell)|=1$. Moreover, it vanishes for all arguments $\ell$  of the class  ${\cal M}_2$. 
If  $\ell$ and  $N$  have no factors in common we find 
$|{\cal A}_N^{(\ell-1)}|=\sqrt{2/\ell}$, $\sqrt{1/\ell}$ or zero, depending on $\ell$ either being an element of ${\cal M}_0$, ${\cal M}_1$ and ${\cal M}_3$, or ${\cal M}_2$, respectively. The corresponding curves are represented by a dashed, a solid or a dashed line, respectively.  
Data points which are not situated on one of these three curves indicate that the corresponding arguments $\ell$ carry information on the factors of $N$. For example, $\ell=12= 4 \cdot 3$ shares the  factor $s=3$ with $N=1911$. Since $k=4$ is an element of ${\cal M}_0$ we find according to \eq{eq:A_factor} the value $\sqrt{1/2}$. 
}
\label{B1911}
\end{figure}

In \fig{B1911} we show as an example the signal $|{\cal A}_N^{(\ell-1)}(\ell)|$ for the number $N=1911=3\cdot 7^2\cdot 13$. The three curves $|{\cal A}_N^{(\ell-1)}(\ell)|=\sqrt{2/\ell},\,\sqrt{1/\ell},\textrm{ and } 0$ are indicated by a dashed, a solid, and a dashed line, respectively. The arguments $\ell$ of data points which are not situated on one of these curves contain information on the factors of $N$.

\section{Conclusions and discussion}
\label{conclusion}
In the present article we have analyzed different schemes to factor numbers based on three classes of Gauss sums. We have developed analytical criteria for deducing the  factors of a number $N$ from a physical signal given by a Gauss sum and have demonstrated  the suggested schemes by numerical examples. The continuous version of Gauss sum factorization  has a remarkable scaling property. As a result, a single realization of the Gauss sum  for the number $N$ yields information on the factors of another number $N'$. The discrete version of this scheme rests solely on the analysis of Gauss sums at integer arguments. The Gauss reciprocity relation allowed us to establish the link with yet another type of Gauss sum. Here, the roles of the argument and the number to be factored are interchanged.

Unlike Shor's algorithm, the factorization schemes discussed in the present article do not feature a reduction of the  computational resources.
Nevertheless, we are convinced that this approach will open new perspectives on the connection between physics and number theory \cite{merkel:2006:a,schleich:2005:primes}. In particular, our work is motivated by the search for  physical systems which reproduce Gauss sums. In this spirit, the results of this article provide the mathematical background for part II where we address  physical realizations of Gauss sums. 

\ack{
We have profited from numerous and fruitful discussions with M Arndt, W~B Case, B Chatel, C Feiler, M Gilowski, D Haase , M~Yu Ivanov, E Lutz, H Maier, M Mehring, A~A Rangelov, E~M Rasel, M Sadgrove, Y Shih, M $\check{\textrm{S}}$tefa\'n$\check{\textrm{a}}$k, D Suter, V Tamma, S Weber,   and M S Zubairy.
W~M and W P S acknowledge financial support by 
the  Baden-W\"{u}rttemberg Stiftung. Moreover, W P S also would 
like to thank the Alexander von Humboldt Stiftung and the 
Max-Planck-Gesellschaft for receiving the Max-Planck-Forschungspreis. I Sh A thanks the Israel Science Foundation for supporting this work.
Our research has also profited immensely from the stimulating atmosphere of the \textit{Ulm Graduate School Mathematical Analysis of Evolution, Information and Complexity} under the leadership of W Arendt. 
}

\appendix
\setcounter{section}{1}
\section{Quantum algorithm for phase of Gauss sum\label{algorithm for Gauss phase}}

In this appendix we briefly summarize a quantum algorithm \cite{vandam1,vandam2}  proposed by W. van Dam and  G. Seroussi, which evaluates the phase of  a certain class of Gauss sums. We compare and contrast these Gauss sums with the ones central to our article. Since the algorithm  relies heavily on properties of multiplicative characters, we also briefly motivate  relevant identities.

\subsection{Gauss sums over rings}

Gauss sums over  a finite ring  consisting of integers modulo $n$ denoted by  $\mathbb{Z}/n\mathbb{Z}$ read 
\be
G(\mathbb{Z}/n\mathbb{Z},\chi,\beta)\equiv \sumlim_{x =0}^{n-1}\; \chi(x)\; \textrm{exp}\left[ 2 \pi i\;\frac{ \beta x}{n}\right]\label{eq:defgauss2} 
\ee
where $\chi=\chi(x)$ is a multiplicative character.

The absolute value of $G$ is given by
\be
\left| G(\mathbb{Z}/n\mathbb{Z},\chi,\beta)\right|=  \sqrt{n}\label{eq:betrag_G}
\ee
provided $n$ is prime. All other cases can be reduced to that one. 

In the present the quadratic Gauss sum
\be
G(a,b,c)\equiv \sumlim_{m=0}^{b-1} \textrm{exp}\left[ 2\pi \rmi (m\frac{c}{b}+m^2\frac{a}{b})\right]
\ee
plays a central role. However, only for $c=0$, gcd($a$,$b$)=1, and b squarefree, which means $b\neq n \cdot q^2$, we can cast it into the form 
\be
G(a,b,0)= \sumlim_{k \epsilon \mathbb{Z}/b\mathbb{Z}}\left(\frac{k}{b}\right)\; \textrm{exp}\left[2\pi i \frac{ka}{b}\right]
\ee
of a Gauss sums over a finite ring, where $\left(\frac{k}{c}\right)$ denotes the Legendre symbol.
Hence, only in this case we can perform the algorithm proposed in Ref.\cite{vandam1,vandam2}.

Furthermore, we note that for the quadratic Gauss sum $G(a,0,c)$, it is very easy to estimate the phase $\gamma$, in contrast to the absolute value, which depends on the greatest common divisor of $a$ and $c$. In Sec.\ref{gauss:factor} we use this attribute of quadratic Gauss sums to factor numbers.

\subsection{Outline of algorithm \label{sec:outline_of_algorithm}}

In order to bring out most clearly the main principles of this phase-determining algorithm \cite{vandam1,vandam2}, we choose the example of a finite ring $\mathbb{Z}/n\mathbb{Z}$ over a prime $n$. If $n$ is not prime, we have to first find the factors of $n$ and then execute the algorithm for each factor of $n$. With the help of the phases corresponding to each factor, it is possible \cite{vandam1,vandam2} to estimate the phase of the Gauss sum. 

Moreover, in the next section we establish the relation 
\be
G(\mathbb{Z}/n\mathbb{Z},\chi,\beta)= \chi^{-1}(\beta)G(\mathbb{Z}/n\mathbb{Z},\chi,1)\label{eq:reduction}
\ee
which shows that 
it suffices to find the phase of  $G(\mathbb{Z}/n\mathbb{Z},\chi,1)$ rather than  $G(\mathbb{Z}/n\mathbb{Z},\chi,\beta)$. 
With the help of Shor's discrete log algorithm \cite{shor:1997} the phase $\gamma_1$ of $\chi^{-1}(\beta)$ can be calculated in an efficient way. Hence, we are left with the task to find the phase $\gamma_2$ of $G(\mathbb{Z}/n\mathbb{Z},\chi,1)$.

The algorithm proceeds in four steps: 
First we prepare the superposition state
\be
\ket{\psi}\equiv\frac{1}{\sqrt{2}}(\ket{\oslash}+\ket{\chi})\label{eq:def_psi}
\ee
consisting of the state
\be
\ket{\chi}\equiv \frac{1}{\sqrt{n-1}}\sum_{x =0}^{n-1} \chi(x) \ket{x},
\ee
and the 'stale' component $\ket{\oslash}$. We emphasize that $\ket{\chi}$ is a superposition of only $n-1$ orthonormalized  states $\ket{x}$ since $\chi(0)=0$ as shown in the next section.

In the next step we perform a Quantum Fourier Transformation (QFT)
\be
\hat U _{\textrm{QFT}} \ket{x}\equiv\frac{1}{\sqrt{n}}\sum_{y=0 }^{n-1}\textrm{exp}\left[2\pi \rmi\frac{xy}{n}\right]\ket{y}
\ee
of $\ket{\chi}$, which leads us to
\be
\hat U _{\textrm{QFT}} \ket{\chi}= \frac{1}{\sqrt{n(n-1)}}\sum_{y=0 }^{n-1}\sum_{x=0 }^{n-1}\chi(x)\;\textrm{exp}\left[2\pi \rmi\frac{xy}{n}\right]\ket{y}.
\ee
With the help of the definition of $G$, \eq{eq:defgauss2}, we arrive at
\be
\hat U_{\textrm{QFT}} \ket{\chi}=\frac{1}{\sqrt{n(n-1)}}\sum_{y=0 }^{n-1}G(\mathbb{Z}/n\mathbb{Z},\chi,y)\ket{y}
\ee
which reduces with the identity \eq{eq:reduction} to
\be
\hat U_{\textrm{QFT}} \ket{\chi}= \frac{G(\mathbb{Z}/n\mathbb{Z},\chi,1)}{\sqrt{n(n-1)}}\sum_{y=0 }^{n-1}\chi^{-1}(y)\ket{y}.
\ee

In the third step we produce the phase shift 
\be
\hat U_p\ket{y} \equiv \chi^2(y)\ket{y}
\ee
on the basis states $\ket{y}$ which yields
\be
\hat U_p\; \hat U_{\textrm{QFT}}\;\ket{\chi} = \frac{G(\mathbb{Z}/n\mathbb{Z},\chi,1)}{\sqrt{n}} \ket{\chi}.
\ee
Here we have used again the definition of $G$, \eq{eq:defgauss2}. 

The 'stale' component $\ket{\oslash}$ is invariant under the phase transformation. As a result the transformed superposition state $\ket{\tilde \psi}=\hat U_p\; \hat U_{\textrm{QFT}}\ket{\psi}$ reads
\be
\ket{\tilde \psi}=\frac{1}{\sqrt{2}}(\ket{\oslash}+\frac{G(\mathbb{Z}/n\mathbb{Z},\chi,1)}{\sqrt{n}}\ket{\chi})
\ee
which due to the expression\eq{eq:betrag_G} for the absolute value of $G$ takes the form
\be
\ket{\tilde \psi}= \frac{1}{\sqrt{2}}(\ket{\oslash}+\textrm{e}^{i\gamma_2}\ket{\chi}),\label{eq:def_psi_tilde}
\ee
where $\gamma_2$ denotes the phase of $G$.

In the fourth and final step we implement, with the help of the initial superposition state $\ket{\psi}$, \eq{eq:def_psi} and the transformed state $\ket{\tilde \psi}$ \eq{eq:def_psi_tilde}
an algorithm to determine the phase $\gamma_2$.

In Ref.\cite{vandam1,vandam2} it is shown, that this algorithm consisting of state preparation, Quantum Fourier Transform, creation of phase shift and phase estimation can be performed in polylogarithmic time.

\subsection{Properties of multiplicative characters\label{sub:character}}

In this section we verify the identity  
\be
G(\mathbb{Z}/n\mathbb{Z},\chi,\beta)= \chi^{-1}(\beta)G(\mathbb{Z}/n\mathbb{Z},\chi,1).\label{eq:reduction2}
\ee
which is crucial to the algorithm discussed in the preceding section. For this purpose we first derive some properties of multiplicative characters and then use them to establish \eq{eq:reduction2}.

We start from the identity 
\be
\chi(xy)= \chi(x)\chi(y)\label{eq:multi}
\ee
of the multiplicative character $\chi$ which for $y\equiv 1$ yields 
\be
\chi(x \cdot 1)= \chi(x)=\chi(x)\chi(1),
\ee
that is 
\be
\chi(1)=1.\label{eq:chi1}
\ee

Likewise, we find from \eq{eq:multi} for $y\equiv 0$ the relation
\be
\chi(x \cdot 0)= \chi(0)=\chi(x)\chi(0)
\ee
which is only true for arbitrary $x$ provided 
\be
\chi(0)=0\label{chi=0}.
\ee

Next, we substitute  $y\equiv x^{-1}$ into \eq{eq:multi} and   arrive at
\be
\chi(1)=\chi(x \frac{1}{x})= \chi(x)\chi(x^{-1})
\ee 
which with the help of \eq{eq:chi1} reduces to
\be
\chi(x^{-1})=\chi^{-1}(x).
\ee
As a consequence, we obtain the relation 
\be
\chi\left( x\beta^{-1}\right) =  \chi\left( x\right)\chi\left( \beta^{-1}\right) = \chi\left( x\right)\chi\left( \beta\right)^{-1}.\label{eq:char}
\ee
We are now in the position to verify the identity \eq{eq:reduction2}. For this purpose we introduce the summation index $y \equiv \beta x$ in the Gauss sum,\eq{eq:defgauss2}, which reads 
\be
G(\mathbb{Z}/n\mathbb{Z},\chi,\beta)=\sumlim_{y =0}^{n-1}\; \chi(y \beta^{-1})\; \textrm{exp}\left[ 2 \pi\rm i\;\frac{  y}{n}\right].
\ee
Here, we emphasize that the summation has not changed due to the modular structure of the domain of arguments $x$ of the Gauss sum.

With the help of \eq{eq:char} and the definition \eq{eq:defgauss2} of $G$ we immediately arrive at \eq{eq:reduction2}.

\section{Factorization with an N-slit interferometer\label{N-slit interferometer}}

In this appendix we summarize the essential ingredients of the pioneering proposal \cite{clauser:1996} by J.F.Clauser and J.P.Dowling to factor odd numbers using a  Young N-slit interferometer. For a more elementary argument we refer to the appendix  of Ref.\cite{mack:2009}.

We operate the interferometer with matter waves whose propagation is  governed by the Schr\"odinger equation rather than light waves, whose evolution is described by the Maxwell equations. Needlees to say, in the paraxial limit both wave equations agree. 

We first derive an expression for the wave function  on a screen located at a distance $z$ from an N-slit grating with the period $d$.  Screen and grating are parallel to each other. We then cast the Green's function into a form which allows us to derive a criterion for the factorization of  an odd integer $N$. Here, we use a property of Gauss sums derived in Appendix \ref{ss:W(c+1/2)}.

\subsection{Wave function on the screen}


For the propagation of the initial wave function 
\be\psi_0(x) \equiv \psi(x,0) \equiv \sum\limits_{n=0}^{N-1}\phi(x-nd)\label{initial wave}
\ee
representing an array of $N$ identical wave functions $\phi=\phi(x)$ separated by $d$
 we start from the Huygens integral \cite{feynman:1965}
\be
\psi(x,t)= {\cal N}(t) \int\limits_{-\infty}^{\infty}\rmd y \: \textrm{e}^{\rmi\alpha(t) (x-y)^2}\psi_0(y)\label{Huygen}
\ee
 for matter waves. Here we have introduced the abbreviation
\be
\alpha(t) \equiv \frac{M}{2 \hbar t}\label{def:alpha}
\ee
together with
\be
 {\cal N} (t) \equiv \sqrt{\frac{\alpha(t)}{\rmi\pi }}
\ee
and $M$ denotes the mass of the particle.

We substitute the initial wave function, \eq{initial wave}, into the Huygens integral, \eq{Huygen}, and introduce the new integration variable $\zeta\equiv y/d-n$ which yields
\be
\psi(x,t)= \int\limits_{-\infty}^{\infty} \rmd\zeta\; \textfrak{G}(\frac{x}{d}-\zeta,t)\phi(d\zeta),\label{eq:huygens}
\ee
where 
\be
\textfrak{G}(\xi,t)\equiv d{\cal N}(t) \sum_{n=0}^{N-1}\textrm{e}^{\rmi\alpha(t) d^2 (\xi-n)^2}\label{Greens function}
\ee
is the Green's function of the $N$-slit problem.

When we can treat  the motion along the z-axis classically the time $t$ translates into the longitudinal position $z$ via the identity
\be t= \frac{z}{v_z}
\ee 
where $v_z$ is the  velocity along the $z$-axis.

With the help of the de Broglie relation $ Mv_z = \hbar 2 \pi /\lambda$ containing the wavelength $\lambda$ of the matter wave together with the definition of $\alpha$, \eq{def:alpha}, we establish the identity
\be
\alpha (t)=\alpha \left( \frac{z}{v_z}\right)  = \frac{M v_z}{2\hbar z} = \frac{\pi}{\lambda z}
\ee
 and the Green's function, \eq{Greens function}, takes the form 
\be
\textfrak{G}\left(\xi,\frac{z}{v_z}\right) = \sqrt{\frac{d^2}{\rmi\lambda z}}\sum_{n=0}^{N-1}\textrm{exp}\left[ \rmi \pi \frac{d^2}{z \lambda} (\xi-n)^2\right]\label{eq:G_N_slit} .
\ee

We conclude by evaluating the Huygens integral, \eq{eq:huygens}, in an approximate way. Whenever $\phi$ is mainly concentrated around $y=0$ and narrow compared to the smallest length scale of $\textfrak{G}$, we can evaluate $\textfrak{G}$ at $y=0$ and factor it out of the integral, which leads us to
\be
\psi(x,\frac{z}{v_z})\approx \bar{\phi}\; \textfrak{G}(\frac{x}{d},\frac{z}{v_z})
\ee
where 
\be
\bar{\phi}\equiv \int\limits_{-\infty}^{\infty}\rmd y\; \phi(y).
\ee
Hence,  the intensity pattern $|\psi\left( x,z/v_z\right) |^2$ on the screen is mainly determined by the Green's function. 

\subsection{Different representation of Green's function}

The factorization property \cite{clauser:1996} of the N-slit interferometer  results from the periodicity properties of the Green's function $\textfrak{G}$, \eq{eq:G_N_slit}. In order to verify this claim we now analyze the dependence of $\textfrak{G}$ on $\lambda$. 

Of particular importance is the situation when $\lambda= l d^2/z$, where $l$ is an integer. In this case $\textfrak{G}$ given by \eq{eq:G_N_slit}  reduces to
\be
\textfrak{G}= \sqrt{\frac{1}{\rmi l }}\sum_{n=0}^{N-1}\textrm{exp}\left[ \rmi \pi \frac{1}{l} (\xi-n)^2\right].
\ee

It is instructive to decompose the complete square in the phase into the individual contributions and cast the sum in the form 
\be
\textfrak{G}= \textrm{e}^{\rmi\Phi} \mathcal{G}
\ee
where 
\be
\Phi \equiv \frac{\pi}{l}\xi^2 -\frac{\pi}{4}
\ee
and
\be
\mathcal{G}\equiv\sqrt{\frac{1}{ l }} \sum_{n=0}^{N-1}\textrm{exp}\left(-2\pi\rmi \frac{n}{l} \xi  \right)\gamma_n
\ee
with
\be
\gamma_n \equiv \textrm{exp}\left(\rmi \pi \frac{n^2}{l}\right).
\ee
Since the phase factor $\gamma_n$ satisfies the periodicity property
\be
\gamma_{n+l}= \textrm{exp}\left(\rmi\pi l\right)\gamma_n\label{symmetry}
\ee
it is useful to represent the number $N$ to be factored
as a multiple $k$ of $l$ plus a remainder $r$ that is $N = k \cdot \ell + r$.

Similarly, the summation index $n$ can be decomposed as $n= jl+p$.
This property suggests to express the sum over $n$ into a double sum where the sum over $j$ covers the $k$ periods of length $l$ and the one over $p$ contains the elements of a single period. In case $l$ is not a factor of $N$, that is when $r$ is non-vanishing, there will be a remainder $\mathcal{R}$ due to the summation over an incomplete unit cell. Indeed, we arrive at
\be
\mathcal{G}= \sqrt{\frac{1}{ l }} \sum_{j=0}^{k-1} \sum_{p=0}^{l-1}\textrm{exp}\left[-2\pi \rmi \frac{jl+p}{l} \xi  \right]\gamma_{jl+p}+\mathcal{R}(\xi)
\ee
with
\be
\mathcal{R}(\xi)\equiv \sqrt{\frac{1}{ l }} \sum_{p=0}^{r-1}\textrm{exp}\left[-2\pi\rmi \frac{kl+p}{l} \xi  \right]\gamma_{kl+p}.
\ee

Due to the periodicity property \eq{symmetry} we find $\gamma_{jl+p}=  \textrm{exp}\left[\rmi\pi  jl\right]\;\gamma_{p}$ and the two sums in $\mathcal{G}$ over $j$ and $p$ decouple. As a result $\mathcal{G}$ takes the form
\be
\mathcal{G}(\xi)= \mathcal{W}^{(l)}(\xi) \Delta_k(\xi-l/2) +\mathcal{R}(\xi)
\ee
where we have introduced the abbreviations
\be
\mathcal{W}^{(l)}(\xi)\equiv \sqrt{\frac{1}{l}} \sum_{p=0}^{l-1}\textrm{exp}\left[   \pi \rmi (p^2\frac{1}{l}-2p\frac{\xi}{l})  \right] \label{eq:clausergauss}
\ee
and
\be
\Delta_k(\zeta)\equiv \sum_{j=0}^{k-1}\textrm{exp}\left[-2\pi \rmi  j\zeta   \right].
\ee

Moreover, the remainder $\mathcal{R}$ reduces to
\be
\mathcal{R}=  \textrm{exp}\left[-2\pi \rmi k (\xi-l/2)\right] \sqrt{\frac{1}{ l }}\sum_{p=0}^{r-1}\textrm{exp}\left[ 2\pi \rmi \left(\frac{p^2}{l}-p\frac{\xi}{l}\right)  \right]. \label{eq:reminder}
\ee

When we compare the sums in ${\cal W}^{(l)}$ and the reminder $\cal R$ given by \eq{eq:clausergauss} and \eq{eq:reminder} we find that they only differ in their upper limits. Indeed, the summation in ${\cal W}^{(l)}$ extends over the period $l$ whereas the one in $\cal R$ only contains $r$ terms.

\subsection{Criterion for factorization}

Since the function $\Delta_k= \Delta_k(\zeta)$ is sharply peaked when $\zeta$ is an integer $s$, the Green's function $\mathcal{G}$ has maxima at positions $\xi = \ell/2 + s$. Because $N$ is odd, $l$ must be also odd in order to be an factor of $N$. Hence, the phases interfere constructively at half integers, that is at $\xi = q + 1/2$.

The intensity  of the peaks is given by $|\mathcal{W}^{(l)}(\xi=q+1/2)|^2$.  
As shown in Appendix \ref{ss:W(c+1/2)} it is equal to unity independent from $q$. 

We therefore arrive at the following criterion \cite{clauser:1996}for finding factors of an odd number N: if $l$ is a factor of  $N$,  the remainder $\mathcal R$ vanishes and the diffraction pattern consists of spikes at the positions $x= (q+1/2)d $ of identical hight. If $l$ is not a factor of $N$ the remainder $\mathcal{R}$
leads to diffraction patterns, which interfere with these spikes. As a result their height is not identical anymore.



\section{Absolute value of a finite Gauss sum\label{ss:W(c+1/2)}}

In the factorization schemes based on the continuous Gauss sum $S_N(\xi)$ and the N-slit interferometer discussed in Sec. \ref{gauss:factor} and Appendix \ref{N-slit interferometer}, respectively, the relation 
\be
|\tilde{\cal W}^{(r)}(a,b,c)|^2=\frac{1}{r}\label{eq:W_tilde_cal}
\ee
for the sum
\be
\tilde{\cal W}^{(r)}(a,b,c)\equiv \frac{1}{r}\sum_{p=0}^{r-1}\exp\left[ \frac{\rmi\pi }{r} \left(p^2a+2bp+pc\right) \right] \label{eq:tilde_W}
\ee
plays a central role.
It is noteworthy that \eq{eq:W_tilde_cal} holds true for any integer $a$, $b$ and $c$ provided $a$ and $r$ are coprime and $ar-c$ even. 
In the present section we verify \eq{eq:W_tilde_cal}.

For this purpose we first note that $\tilde{\cal W}^{(r)}$ is periodic in $b$ with period $r$. As a result we can expand $|\tilde{\cal W}^{(r)}|^2$ into a Fourier series 
\be
\left|\tilde{\cal W}^{(r)}(a,b,c)\right|^2= \frac{1}{r}\sum_{n=0}^{r-1} A^{(r)}(n) \exp\left[-2\pi \rmi \frac{b}{r}n\right]\label{eq:fourier_rep}
\ee
with the expansion coefficients
\be
A^{(r)}(n) \equiv \sum_{b=0}^{r-1}\left|\tilde{\cal W}^{(r)}(a,b,c)\right|^2 \textrm{exp}\left[2\pi \rmi\frac{n}{r}b\right] .
\ee
In the second step, we  substitute the definition \eq{eq:tilde_W} of $\tilde{\cal W}^{(r)}$ into this expression for $A^{(r)}(n)$ and find
\begin{eqnarray}
A^{(r)}(n) =& \frac{1}{r^2}\sum_{p,p'=0}^{r-1} \textrm{exp}\left[\frac{\rmi\pi}{r}\left(p^2a-p'^2a+pc-p'c\right)\right]
\Delta(p-p'+n). \label{eq:fourier}
\end{eqnarray}
Here, all terms containing the variable $b$ are combined in 
\be
\Delta(s) \equiv \sum_{b=0}^{r-1} \textrm{exp}\left[2\pi \rmi\frac{s}{r} b\right]
= \left\{
\begin{array}{cl}r&\textrm{for }s= k\cdot r \\ 0 & \textrm{else}\end{array}\right.\nonumber.
\ee
Because $0 \leq p,p',n < r$, there are only two possibilities for which $\Delta(p-p'+n)$ is non-vanishing: $p-p'+n = 0$ or $p-p'+n = r$. In the first case, $p$ must be smaller than $ r-n$ because $p'<r$. In the second case $p$ must be larger or equal to $r-n$ because $0\leq p'$.  Therefore,   the Fourier coefficients given by \eq{eq:fourier} reduce to
\begin{eqnarray}
\fl A^{(r)}(n) = \frac{1}{r}\sum_{p=0}^{r-n-1}\textrm{exp}\left[-\rmi\frac{\pi }{r}\left(2pna+an^2+nc\right)\right]\nonumber\\
+ \frac{1}{r}\sum_{p=r-n}^{r-1}\textrm{exp}\left[-\rmi\frac{\pi }{r}\left(2pna+an^2+nc\right)\right]\textrm{e}^{-\pi i (ar-c)}
\end{eqnarray}
If $ar-c$ is an even number,  we can combine these two sums to one sum
\be
A^{(r)}(n) = \frac{1}{r}\sum_{p=0}^{r-1}\exp\left[-2\pi \rmi \frac{na}{r}p\right]\exp\left[-\frac{\rmi\pi}{r}(an^2+nc)\right]\label{eq:fourier_fast_ende}
\ee
which yields
\be
A^{(r)}(n)=\left\{\begin{array}{cl}1&\textrm{for }n= 0\\ 0 & \textrm{else}\end{array}\right..\label{eq:fourier_ende}
\ee
In the last step we have used the fact, that $a$ and $r$ are coprime.

When we now substitute \eq{eq:fourier_ende} into the Fourier representation  \eq{eq:fourier_rep} we immediately arrive at \eq{eq:W_tilde_cal}.

We conclude by analyzing the  finite Gauss sum 
\be
{\cal W}^{(r)}_m\equiv\frac{1}{r}\sumlim_{p=0}^{r-1}
\exp\left[2\pi \rmi\,\left(p^2\frac{q}{r}+p\frac{m}{r}\right)\right]\label{eq:finite_gauss_2}
\ee
which is a special case of $ \tilde{\cal W}^{(r)}$. Indeed, a comparison between \eq{eq:tilde_W} and \eq{eq:finite_gauss_2} allows us to identify the parameters $a=2q$, $b=m$ and $c=0$. However, the discussion is now slightly more complicated. If $r$ is odd it cannot share a divisor with $a=2q$ and we can apply \eq{eq:W_tilde_cal}. In this case only the Fourier term with $n=0$ is non-vanishing. On the other hand, if $r$ is even, it shares with $a=2q$ the  factor $2$ and   also the term

\begin{eqnarray}
A^{(r)}(n=r/2)=&\frac{1}{r} \sum\limits_{p=0}^{r-1}\exp\left[-2\pi\rm i qp-\rmi \pi \frac{qr}{2}\right] \nonumber\\
=&\exp\left[-\rmi \pi \frac{qr}{2}\right]
\end{eqnarray}

is non-vanisching. With the help of \eq{eq:fourier_rep} we find
\be
\left|\tilde{\cal W}^{(r)}(2q,m,0)\right|^2= \frac{1}{r}\left(1+\exp\left[-\rmi \pi \left(\frac{qr}{2}+m\right)\right]\right)
\ee
can either be zero if $qr/2+m$ is odd, or $2/r$ if it is even.




\section*{References}
\bibliographystyle{bst}


\end{document}